\documentclass[aps,prd,reprint,twocolumn,superscriptaddress]{revtex4-1}
\usepackage[english]{babel}
\usepackage[T1]{fontenc}
\usepackage{amsmath}
\usepackage{amssymb}
\usepackage{graphicx}
\usepackage{color}
\usepackage{times}
\usepackage{euscript}
\usepackage{amsthm}
\usepackage{amsfonts}
\usepackage[english]{babel}
\usepackage{epstopdf}
\usepackage{multirow,makecell,array}
\usepackage{mathtools}
\usepackage{float}
\usepackage[dvipsnames]{xcolor}
\usepackage{graphicx}
\usepackage{bbm}

\newcommand*{\I}{i}

\begin{document}

\title{Kinetic theory of vacuum pair production in uniform electric fields revisited}

\author{I.~A.~Aleksandrov}
\email{i.aleksandrov@spbu.ru}
\affiliation{Department of Physics, Saint Petersburg State University, Universitetskaya Naberezhnaya 7/9, Saint Petersburg 199034, Russia}
\affiliation{Ioffe Institute, Politekhnicheskaya Street 26, Saint Petersburg 194021, Russia}
\author{A.~Kudlis}
\email{andrewkudlis@gmail.com}
\affiliation{Science Institute, University of Iceland, Dunhagi 3, IS-107, Reykjavik, Iceland}
\author{A.~I.~Klochai}
\affiliation{Department of Physics, Saint Petersburg State University, Universitetskaya Naberezhnaya 7/9, Saint Petersburg 199034, Russia}

\begin{abstract}
We investigate the phenomenon of electron-positron pair production from vacuum in the presence of a uniform time-dependent electric field of arbitrary polarization. Taking into account the interaction with the external classical background in a nonperturbative manner, we quantize the electron-positron field and derive a system of ten quantum kinetic equations (QKEs) showing that the previously-used QKEs are incorrect once the external field rotates in space. We employ then the Wigner-function formalism of the field quantization and establish a direct connection between the Dirac-Heisenberg-Wigner (DHW) approach to investigating the vacuum pair-production process and the QKEs. We provide a self-contained description of the two theoretical frameworks rigorously proving their equivalence and present an exact one-to-one correspondence between the kinetic functions involved within the two techniques. Special focus is placed on the analysis of the spin effects in the final particle distributions.
\end{abstract}

\maketitle

%%%%%%%%%%%%%%%%%%%%%%%%%%%%%%%%%%%%%%%%%%%%%%%%%%%%%%%%%

\section{Introduction}
\label{sec:intro}

One of the fundamental differences between ordinary quantum mechanics and quantum field theory is that the latter enables elementary particles to participate in reactions, in which the number of any species may not be conserved. In the context of quantum electrodynamics (QED), even in vacuum, this leads to spontaneous acts of creation and annihilation of virtual electron-positron pairs that affect the physical properties of real particles and the corresponding compound systems. Furthermore, these virtual quanta can be directly transformed into real $e^+e^-$ pairs by making the physical vacuum interact with an external electromagnetic field. In the case of a static uniform electric background, this phenomenon is analogous to the process of quantum mechanical tunneling and is known as the Sauter-Schwinger mechanism~\cite{sauter_1931,heisenberg_euler,weisskopf,schwinger_1951}. This fundamental effect is inherently nonperturbative with respect to the external field strength. Within the realm of QED, a particularly exciting challenge revolves around the practical observation of the Sauter-Schwinger pair-production process, which is yet to be experimentally investigated (for recent developments regarding strong-field QED see reviews~\cite{dittrich_gies,dunne_shifman,dipiazza_rmp_2012,xie_review_2017,gonoskov_2022,fedotov_review}).

The mechanism of $e^+e^-$ pair production in realistic setups should be described by taking into account the corresponding spatiotemporal structure of the external fields. For instance, an inhomogeneous electromagnetic background efficiently generating pairs and preserving the nonperturbative nature of the process can be formed in a combination of high-intensity laser pulses. From a theoretical point of view, it is crucial to incorporate the interaction with the external classical field without relying on perturbative techniques. It turns out that an exact analytical treatment of the problem is only possible for a very limited number of simple field configurations, which is basically due to the lack of the analytical solutions to the Dirac equation in the presence of the external background (for exactly solvable setups see Refs.~\cite{fradkin_gitman_shvartsman,gavrilov_prd_1996,adorno_2017,breev_prd_2021} and references therein). Therefore, it is highly desirable to develop nonperturbative numerical methods. Multidimensional inhomogeneities of the external classical field can be exactly taken into account, generally speaking, by means of two approaches. First, one can extract the pair production probabilities and momentum distributions from the complete set of the one-particle solutions of the Dirac equation according to the general formalism of Furry-picture quantization with unstable vacuum~\cite{fradkin_gitman_shvartsman}. This general approach with various modifications has been implemented in numerous studies (see, e.g., Refs.~\cite{gavrilov_prd_1996,adorno_2017,breev_prd_2021,ruf_prl_2009,woellert_prd_2015,aleksandrov_prd_2016,aleksandrov_prd_2017_1,aleksandrov_prd_2017,lv_pra_2018,lv_prl_2018,aleksandrov_prd_2018,aleksandrov_kohlfuerst,sevostyanov_prd_2021}). Second, one can employ the Wigner-function formalism~\cite{vasak_1987,BB_prd_1991,zhuang_1996,zhuang_prd_1998,ochs_1998,hebenstreit_prd_2010,fauth_prd_2021} and describe pair production by solving a system of the Dirac-Heisenberg-Wigner (DHW) equations, which was done, e.g., in Refs.~\cite{aleksandrov_kohlfuerst,hebenstreit_prl_2009,blinne_gies_2014,blinne_strobel_2016,olugh_prd_2019,kohlfuerst_prd_2019,li_prd_2019,kohlfuerst_prd_2020,kohlfuerst_arxiv_2022_2,kohlfuerst_arxiv_2022,yu_prd_2023,hu_prd_2023,hu_arxiv_2024}. If the external field is homogeneous in space, i.e., it depends solely on time in a given reference frame, then one can also utilize the so-called quantum kinetic equations (QKEs), which provide a third computational tool for analyzing vacuum pair production~\cite{GMM,pervushin_2005,pervushin_skokov,schmidt_1998,kluger_prd_1998,schmidt_prd_1999,bloch_prd_1999,dumlu_prd_2009,fedotov_prd_2011,blaschke_prd_2011,aleksandrov_epjst_2020} (see also Refs.~\cite{sevostyanov_prd_2021,alkofer_prl_2001,otto_plb_2015,panferov_epjd_2016,aleksandrov_symmetry,aleksandrov_sevostyanov_2022}).

The above mentioned techniques allows one to exactly treat nonstationary classical backgrounds. Many important analytical insights and quantitative predictions can also be obtained by means of semiclassical approaches (see, e.g., Refs.~\cite{brezin_1970,popov_1972,affleck_1982,popov_2005,dunne_schubert_2005,dumlu_dunne_2011,taya_2021,kohlfuerst_prl_2022,esposti_2023} and references therein). Here we point out that by constructing semiclassical one-particle solutions of the Dirac equation, one can combine the corresponding Wentzel–Kramers–Brillouin method (taking into account the Stokes phenomenon, which is crucial for describing pair creation) with the general Furry-picture quantization formalism in order to extract the pair-production probabilities and momentum spectra of the particles.

The primary aim of the present study is to demonstrate the equivalence of the three exact approaches mentioned above in the case of a time-dependent external electric field of arbitrary polarization. Most important, we will establish an explicit connection between the QKE components and the DHW functions, which was previously done only in the case of a linearly polarized external field~\cite{BB_prd_1991,hebenstreit_prd_2010}.

Another important aspect which will be discussed in this paper concerns the correct form of the QKE system. As will be demonstrated below, the QKEs involve 10 functions instead of 12 as was believed until now (see, e.g., Refs.~\cite{pervushin_skokov,blaschke_prd_2011,aleksandrov_epjst_2020}) and the QKE system has a different form. We will revise the theoretical treatment and derive the correct kinetic equations.

Finally, we will consider the spin anisotropy of the particles produced. To properly address experimentally relevant setups, we will suggest that the electron and positron states are classified according to the particle's helicity. The corresponding number densities will be deduced within the QKE approach by means of the Bogoliubov transformation and also expressed in terms of the DHW functions. Besides, we will explicitly show how to construct the corresponding projection of the Wigner function and to obtain the necessary distributions directly via the DHW formalism.

The paper has the following structure. In Sec.~\ref{sec:furry}, we first discuss a general procedure of the Furry-picture quantization taking into account an external classical field. Second, we introduce the adiabatic set of the Hamiltonian eigenfunctions and present a detailed derivation of the QKEs and their main properties. In Sec.~\ref{sec:dhw}, we examine the DHW formalism. Providing first a brief description of this framework in a general four-dimensional case, we turn to the analysis of the purely time-dependent setups. Here we reduce the DHW system to ten equations and explicitly show how the corresponding functions are expressed in terms of the ten QKE components. Finally, we discuss the physical interpretation of the spin states and provide the particle densities in terms of the helicity quantum number. We conclude in Sec.~\ref{sec:conclusion}.

Throughout the paper, we use the units $\hbar = c = 1$ and assume $e<0$.

%%%%%%%%%%%%%%%%%%%%%%%%%%%%%%%%%%%%%%%%%%%%%%%%%%%%%%%%%

\section{Furry-picture quantization and kinetic theory} \label{sec:furry}

We will first discuss a general approach to canonical quantization of the electron-positron field in the presence of an external background. Although we neglect the quantized part of the electromagnetic field, the classical part is treated nonperturbatively. We will briefly recap a transition from the Schr\"odinger representation to the interaction picture and describe how to extract the electron-positron momentum distributions from the one-particle solutions of the Dirac equation. Next, we will introduce the adiabatic set of the Hamiltonian eigenfunctions in the case of a uniform background and derive the QKEs. This technique will be then related to the DHW formalism in Sec.~\ref{sec:dhw}.

%%%

\subsection{General quantization procedure. Heisenberg representation}

We assume that the external electromagnetic field is described by the $c$-numbered function $A^\mu (x)$, where $x = (t, \, \mathbf{x})$, and vanishes outside the time interval $t_\text{in} < t < t_\text{out}$, where one implies $t_\text{in/out} \to \mp \infty$. We start from the Schr\"odinger picture and, following Ref.~\cite{fradkin_gitman_shvartsman}, introduce two sets of the one-particle Hamiltonian eigenfunctions at $t=t_\text{in}$ and $t=t_\text{out}$, respectively:
\begin{equation}
\begin{aligned}
\mathcal{H}_e (t_\text{in})\, {}_\pm \varphi_n (\mathbf{x}) &= {}_\pm \varepsilon_n (t_\text{in})\, {}_\pm \varphi_n (\mathbf{x}) \,,\\
\mathcal{H}_e (t_\text{out})\, {}^\pm \varphi_n (\mathbf{x}) &= {}^\pm \varepsilon_n (t_\text{out}) \, {}^\pm \varphi_n (\mathbf{x}) \,,
\end{aligned}
\label{eq:in_schroedinger}
\end{equation}
where $\mathcal{H}_e (t) = \boldsymbol{\alpha} [-\I \boldsymbol{\nabla} - e \mathbf{A} (x)] + eA_0 (x) + \beta m$ depends on time due to the presence of the external field and the eigenvalues denoted by plus (minus) are positive (negative). These sets are orthonormal and complete with respect to the usual inner product. The index $n$ incorporates the necessary quantum numbers, e.g., momentum and spin. Let us then expand the field operator according to
\begin{eqnarray}
\begin{aligned}
\psi (\mathbf{x}) &= \sum_{n} \big [\mathrm{a}_n(t_\text{in}) \, {}_+ \varphi_n (\mathbf{x}) + \mathrm{b}^\dagger_n(t_\text{in}) \, {}_- \varphi_n (\mathbf{x})\big ] \,,\\
\psi (\mathbf{x}) &= \sum_{n} \big [\mathrm{a}_n(t_\text{out}) \, {}^+ \varphi_n (\mathbf{x}) + \mathrm{b}^\dagger_n(t_\text{out}) \, {}^- \varphi_n (\mathbf{x})\big ] \,,
\end{aligned}
\label{eq:field_decomposition_schroedinger}
\end{eqnarray}
where we have introduced the electron (positron) creation and annihilation operators $\mathrm{a}^\dagger_n$ ($\mathrm{b}^\dagger_n$) and $\mathrm{a}_n$ ($\mathrm{b}_n$), respectively. These operators obey the usual anticommutation relations. Accordingly, the Hamiltonian
\begin{equation}
H_e (t)= \int \! \psi^\dagger (\mathbf{x}) \mathcal{H}_e (t) \psi (\mathbf{x}) d\mathbf{x}
\end{equation}
is diagonalized at time instants $t_\text{in}$ and $t_\text{out}$. The corresponding vacuum states will be denoted by $|0,t_\text{in} \rangle$ and $|0,t_\text{out} \rangle$, respectively.

To construct the field operators in the Heisenberg representation, one has to introduce the following unitary evolution operator:
\begin{equation}
U_e (t, t') = T \, \mathrm{exp} \Bigg [ -\I \int \limits_{t'}^t H_e (\tau) d\tau \Bigg ] \,,
\label{eq:evolution_operator_e}
\end{equation}
where $T$ stands for the conventional time-ordering operator. The operator~\eqref{eq:evolution_operator_e} is a solution of the equation
\begin{equation}
\I \partial_t U_e (t, t') = H_e(t) U_e (t, t')
\label{eq:evolution_operator_equation}
\end{equation}
with the initial condition $U_e (t, t) = \mathbbm{1}$. We perform a transformation by means of the operator $U_e (0, t)$, so the field gains a temporal dependence: 
\begin{equation}
\psi(t,\mathbf{x}) \equiv \psi(x) = U_e (0, t) \psi(\mathbf{x}) U^\dagger_e (0, t) \,.
\label{eq:psi_heis}
\end{equation}
In the Heisenberg representation, the creation and annihilation operators are then given by
\begin{eqnarray}
\begin{aligned}
a_n(\text{in}) &= U_e (0, t_\text{in}) \mathrm{a}_n (t_\text{in}) U^\dagger_e (0, t_\text{in}) \,,\\
a_n(\text{out}) &= U_e (0, t_\text{out}) \mathrm{a}_n(t_\text{out}) U^\dagger_e (0, t_\text{out}) \,.
\end{aligned}
\label{eq:in_out_operators}
\end{eqnarray}
The other creation and annihilation operators are defined similarly. The unitary transformation preserves the anticommutation relations that take place in the Schr\"odinger picture. The {\it in} and {\it out} vacuum states are defined via
\begin{equation}
\begin{aligned}
|0,\text{in} \rangle &= U_e (0, t_\text{in}) |0,t_\text{in} \rangle \,, \\
|0,\text{out} \rangle &= U_e (0, t_\text{out}) |0,t_\text{out} \rangle \,.
\end{aligned}
\label{eq:heisenberg_vacuum}
\end{equation}
By computing the time derivative of~\eqref{eq:psi_heis} and using
\begin{equation}
\big [\psi(\mathbf{x}), \, H_e (t)] = \mathcal{H}_e(t) \psi(\mathbf{x}) \,,
\end{equation}
one finds that the Heisenberg field operator satisfies the equation
\begin{equation}
\I \partial_t \psi(x) = \mathcal{H}_e(t) \psi(x) \,.
\label{eq:psi_heis_equation}
\end{equation}
This means that the temporal evolution of $\psi(x)$ is determined by the same equation as that for the time-dependent one-particle solutions. Let us introduce the corresponding evolution operator
\begin{equation}
G_e (t, t') = T \, \mathrm{exp} \Bigg [ -\I \int \limits_{t'}^t \mathcal{H}_e (\tau) d\tau \Bigg ] \,.
\label{eq:G_operator}
\end{equation}

Suppose that the Schr\"odinger field operator $\psi (\mathbf{x})$ is known, the Heisenberg operator $\psi (t_\text{out}, \mathbf{x})$ can be constructed in two different ways. First, one can carry out the transformation~(\ref{eq:psi_heis}) at $t=t_\text{out}$. Alternatively, one can propagate the Heisenberg operator $\psi (t_\text{in}, \mathbf{x})$ by utilizing the operator $G_e (t_\text{out}, t_\text{in})$. This can be expressed by the following relations:
\begin{eqnarray}
\begin{aligned}
\psi (t_\text{out}, \mathbf{x}) &= U_e (0, t_\text{out}) \psi(\mathbf{x}) U^\dagger_e (0, t_\text{out}) \,,\\
\psi (t_\text{out}, \mathbf{x}) &= G_e (t_\text{out}, t_\text{in}) U_e (0, t_\text{in}) \psi(\mathbf{x}) U^\dagger_e (0, t_\text{in}) \,.
\end{aligned}
\label{eq:psi_G}
\end{eqnarray}
These allow one to express the {\it out} operators in terms of the {\it in} operators with the aid of the $G$-matrix elements. Using Eqs.~\eqref{eq:field_decomposition_schroedinger} and \eqref{eq:in_out_operators}, and the fact that the stationary states ${}_\pm \varphi_n (\mathbf{x})$ and ${}^\pm \varphi_n (\mathbf{x})$ form orthonormal and complete sets, one obtains
\begin{eqnarray}
a_n (\text{out}) &=& \sum_k \big [ a_k (\text{in}) G({}^+|{}_+)_{nk} + b^\dagger_k (\text{in}) G({}^+|{}_-)_{nk} \big ] \,, \label{eq:out_in_1}\\
b_n (\text{out}) &=& \sum_k \big [ a^\dagger_k (\text{in}) G({}_+|{}^-)_{kn} + b_k (\text{in}) G({}_-|{}^-)_{kn} \big ] \,, \label{eq:out_in_2}
\end{eqnarray}
where
\begin{equation}
\begin{aligned}
&G({}^\zeta|{}_\kappa)_{nk} = \int \! d\mathbf{x} \, {}^\zeta \varphi^\dagger_n (\mathbf{x}) G_e (t_\text{out}, t_\text{in})\, {}_\kappa \varphi_k (\mathbf{x}) \,,\\
&G({}_\zeta|{}^\kappa)_{nk} = \big [ G({}^\kappa|{}_\zeta)_{kn} \big ]^* \,, \quad \zeta, \kappa = \pm \,.
\end{aligned}
\label{eq:G_matrix_def}
\end{equation}
Similarly one finds
\begin{eqnarray*}
a_n (\text{in}) &=& \sum_k \big [ a_k (\text{out}) G({}_+|{}^+)_{nk} + b^\dagger_k (\text{out}) G({}_+|{}^-)_{nk} \big ] \,,\\
b_n (\text{in}) &=& \sum_k \big [ a^\dagger_k (\text{out}) G({}^+|{}_-)_{kn} + b_k (\text{out}) G({}^-|{}_-)_{kn} \big ] \,.
\end{eqnarray*}
The $G$ matrices are also orthonormal and complete in the following sense:
\begin{eqnarray}
\sum_\zeta G({}^\pm|{}_\zeta) G({}_\zeta|{}^\mp) &=& \sum_\zeta G({}_\pm|{}^\zeta) G({}^\zeta|{}_\mp) = 0 \,, \label{eq:G_matrix_orthonorm}\\
\sum_\zeta G({}^\pm|{}_\zeta) G({}_\zeta|{}^\pm) &=& \sum_\zeta G({}_\pm|{}^\zeta) G({}^\zeta|{}_\pm) = \mathrm{I} \,. \label{eq:G_matrix_complete}
\end{eqnarray}
The Heisenberg field operator $\psi (x)$ satisfies Eq.~\eqref{eq:psi_heis_equation} and thus evolves in time according to
\begin{equation}
\psi (x) = G_e(x^0, t_\text{in}) \psi (t_\text{in}, \mathbf{x}) \,, \label{eq:psi_G_evolution}
\end{equation}
where the initial field operator $\psi (t_\text{in}, \boldsymbol{x})$ is given by
\begin{equation}
\psi (t_\text{in}, \mathbf{x}) = \sum_{n} \big [a_n(\text{in}) \, {}_+ \varphi_n (\mathbf{x}) + b^\dagger_n(\text{in}) \, {}_- \varphi_n (\mathbf{x})\big ] \,. \label{eq:psi_initial}
\end{equation}
Here we have employed Eqs.~\eqref{eq:field_decomposition_schroedinger}, \eqref{eq:psi_heis}, and \eqref{eq:in_out_operators}. Since the operator $G_e$ evolves not only the field operator, but also the one-particle solutions, it follows from Eqs.~\eqref{eq:psi_G_evolution} and \eqref{eq:psi_initial} that
\begin{equation}
\psi (x) = \sum_{n} \big [a_n(\text{in}) \, {}_+ \varphi_n (x) + b^\dagger_n(\text{in}) \, {}_- \varphi_n (x)\big ] \,, \label{eq:psi_x_in}
\end{equation}
where ${}_+ \varphi_n (x)$ and ${}_- \varphi_n (x)$ are the so-called {\it in} solutions of the Dirac equation which coincide with the eigenfunctions defined in Eqs.~\eqref{eq:in_schroedinger}, i.e. ${}_\zeta \varphi_n (x) = G_e(x^0, t_\text{in}) {}_\zeta \varphi_n (\mathbf{x})$. Note that one was also able to introduce first the {\it in} solutions, decompose the Heisenberg field operator according to Eq.~\eqref{eq:psi_x_in}, and then show that the corresponding operators $a_n(\text{in})$ and $b^\dagger_n(\text{in})$ are indeed time-independent and can be obtained via Eqs.~\eqref{eq:in_out_operators}.

The same way, one finds
\begin{equation}
\psi (x) = \sum_{n} \big [a_n(\text{out}) \, {}^+ \varphi_n (x) + b^\dagger_n(\text{out}) \, {}^- \varphi_n (x)\big ] \,, \label{eq:psi_x_out}
\end{equation}
where ${}^+ \varphi_n (x)$ and ${}^- \varphi_n (x)$ are the {\it out} solutions. Obviously, the Hamiltonian in the Heisenberg representation reads
\begin{eqnarray}
H_{e,\text{H}} (t) &=& U_e (0, t) H_e (t) U^\dagger_e (0, t) \nonumber \\
{}&=& \int \! \psi^\dagger (x) \mathcal{H}_e (t) \psi (x) d\mathbf{x} \,. \label{eq:H_heisenberg}
\end{eqnarray}
It has a diagonal form in terms of the {\it in} ({\it out}) particle-number operators only at $t= t_\text{in}$ ($t= t_\text{out}$) because the time-dependent {\it in} ({\it out}) solutions are not in general the eigenfunctions of the one-particle Hamiltonian $\mathcal{H}_e (t)$ at given $t\in (t_\text{in}, \, t_\text{out})$. The {\it adiabatic} one-particle eigenfunctions of $\mathcal{H}_e (t)$ will be utilized in Sec.~\ref{sec:adiabatic}. The $G$ matrices defined in Eqs.~\eqref{eq:G_matrix_def} can also be evaluated via
\begin{equation}
G({}_\zeta|{}^\kappa)_{nk} = ({}_\zeta \varphi_n, {}^\kappa \varphi_k) \,, \quad G({}^\zeta|{}_\kappa)_{nk} = ({}^\zeta \varphi_n, {}_\kappa \varphi_k) \,,
\label{eq:G_inner_product}
\end{equation}
where $ \zeta$, $\kappa = \pm$. Note that these inner products do not depend on time.

Let us now compute the number density of the electrons produced in some given state~$m$ (we use the term ``density'' keeping in mind that $m$ corresponds to a state which belongs to the continuous spectrum). In the Schr\"odinger representation, the calculation is straightforward:
\begin{equation*}
n^{(e^-)}_m = \langle 0, t_\text{in} | U^\dagger_e (t_\text{out}, t_\text{in}) \mathrm{a}^\dagger_m (t_\text{out}) \mathrm{a}_m (t_\text{out}) U_e (t_\text{out}, t_\text{in}) | 0, t_\text{in} \rangle .
\end{equation*}
Using then the relation $U_e (t_\text{out}, t_\text{in}) = U_e (t_\text{out}, 0) U_e (0, t_\text{in})$ and inserting the identity operator $\mathbbm{1} = U_e (t_\text{out}, 0) U^\dagger_e (t_\text{out}, 0)$ between $a^\dagger_m (t_\text{out})$ and $a_m (t_\text{out})$, one obtains
\begin{equation}
n^{(e^-)}_m = \langle 0,\text{in} | a^\dagger_m (\text{out}) a_m (\text{out}) |0,\text{in}\rangle \,.
\label{eq:number_density_heis}
\end{equation}
Once the vacuum persistence amplitude $\langle 0,\text{out} | 0,\text{in}\rangle$ has an absolute value less than unity, the vacuum state may decay via the production of $e^+e^-$ pairs. The electron number density~\eqref{eq:number_density_heis} can be explicitly evaluated by means of Eqs.~\eqref{eq:out_in_1} and \eqref{eq:out_in_2}:
\begin{eqnarray}
n^{(e^-)}_m &=& \sum_n G({}^+|{}_-)_{mn} G({}_-|{}^+)_{nm} \nonumber \\
{}&=& \{ G({}^+|{}_-) G({}_-|{}^+) \}_{mm} \,.
\label{eq:number_density_zeroth_order_G}
\end{eqnarray}
The analogous expressions for the number density of positrons are given by
\begin{eqnarray}
n^{(e^+)}_m &=& \langle 0,\text{in} | b^\dagger_m (\text{out}) b_m (\text{out}) |0,\text{in}\rangle \,,\\ \label{eq:number_density_heis_pos}
n^{(e^+)}_m &=& \sum_n G({}^-|{}_+)_{mn} G({}_+|{}^-)_{nm} \nonumber \\
{}&=& \{ G({}^-|{}_+) G({}_+|{}^-) \}_{mm} \,. \label{eq:number_density_zeroth_order_G_pos}
\end{eqnarray}
We have demonstrated that the particle number densities can be directly calculated once the one-particle solutions to the Dirac equation in the external field are constructed. It is necessary to find two sets ({\it in} and {\it out}) of the wave functions and compute the $G$ matrices, which represent the corresponding one-particle transition amplitudes. Note that the pair-production probabilities and momentum spectra of the particles produced are encoded in the $G$ matrices involving opposite signs, which is in accordance with the Dirac-sea interpretation.

%%%

\subsection{Uniform external field. Adiabatic basis} \label{sec:adiabatic}

Consider an external time-dependent electric field $\mathbf{E} (t)$ of an arbitrary direction (polarization). We choose the temporal gauge $A_0 = 0$, where $\mathbf{E} (t) = - \mathbf{A}' (t)$. The vector potential at $t = t_\text{in} \to -\infty$ can differ from its limit for $t = t_\text{out} \to +\infty$, so we introduce $\mathbf{A}_\text{in} \equiv \mathbf{A} (t_\text{in})$ and $\mathbf{A}_\text{out} \equiv \mathbf{A} (t_\text{out})$. In practical calculations, it is convenient to choose $\mathbf{A}_\text{out} = \mathbf{0}$ since in the case of a zero vector potential one does not need to distinguish between the canonical momentum and kinetic one and we are interested in computing the momentum distributions of the particles produced at $t = t_\text{out} \to +\infty$.

As the external field is spatially homogeneous, the coordinate dependence of the wave functions is trivial. We specify each solution by the corresponding generalized momentum $\mathbf{p}$ and spin quantum number $s$ and represent the {\it in} and {\it out} functions in the following way:
\begin{eqnarray}
{}_+ \varphi_{\mathbf{p},s} (x) &=& (2\pi)^{-3/2} \, \mathrm{e}^{\I\mathbf{p} \mathbf{x}} \, {}_+ \chi_{\mathbf{p},s} (t) \,, \label{eq:phi_plus_chi_1} \\
{}^+ \varphi_{\mathbf{p},s} (x) &=& (2\pi)^{-3/2} \, \mathrm{e}^{\I\mathbf{p} \mathbf{x}} \, {}^+ \chi_{\mathbf{p},s} (t) \,,\label{eq:phi_plus_chi_2}\\
{}_- \varphi_{\mathbf{p},s} (x) &=& (2\pi)^{-3/2} \, \mathrm{e}^{-\I\mathbf{p} \mathbf{x}} \, {}_- \chi_{\mathbf{p},s} (t) \,, \label{eq:phi_minus_chi_1} \\
{}^- \varphi_{\mathbf{p},s} (x) &=& (2\pi)^{-3/2} \, \mathrm{e}^{-\I\mathbf{p} \mathbf{x}} \, {}^- \chi_{\mathbf{p},s} (t) \,. \label{eq:phi_minus_chi_2}
\end{eqnarray}
The time-dependent parts $\chi$ satisfy the following conditions:
\begin{eqnarray}
{}_+ \chi_{\mathbf{p},s} (t\leqslant t_\text{in}) &=& \mathrm{e}^{-\I p^0(\mathbf{p}-e\mathbf{A}_\text{in}) (t-t_\text{in})} \, u_{\mathbf{p}-e\mathbf{A}_\text{in},s} \,, \label{eq:chi_initial_cond_in_plus} \\
{}^+ \chi_{\mathbf{p},s} (t\geqslant t_\text{out}) &=& \mathrm{e}^{-\I p^0 (\mathbf{p}-e\mathbf{A}_\text{out}) (t-t_\text{out})} \, u_{\mathbf{p}-e\mathbf{A}_\text{out},s} \,, \label{eq:chi_initial_cond_out_plus} \\
{}_- \chi_{\mathbf{p},s} (t\leqslant t_\text{in}) &=& \mathrm{e}^{\I p^0(\mathbf{p}+e\mathbf{A}_\text{in}) (t-t_\text{in})} \, v_{-\mathbf{p}-e\mathbf{A}_\text{in},s} \,, \label{eq:chi_initial_cond_in_minus} \\
{}^- \chi_{\mathbf{p},s} (t\geqslant t_\text{out}) &=& \mathrm{e}^{\I p^0 (\mathbf{p} + e\mathbf{A}_\text{out}) (t-t_\text{out})} \, v_{-\mathbf{p}-e\mathbf{A}_\text{out},s} \,, \label{eq:chi_initial_cond_out_minus}
\end{eqnarray}
where $p^0 (\mathbf{p}) = p_0 (\mathbf{p}) = \sqrt{m^2 + \mathbf{p}^2}$. The bispinors obey
\begin{eqnarray}
\big ( \boldsymbol{\alpha} \mathbf{p} + \beta m \big ) u_{\mathbf{p},s} &=& p^0 (\mathbf{p}) u_{\mathbf{p},s} \,, \label{eq:bispinors_eqs_u} \\
\big ( \boldsymbol{\alpha} \mathbf{p} + \beta m \big ) v_{\mathbf{p},s} &=& -p^0 (\mathbf{p}) v_{\mathbf{p},s} \label{eq:bispinors_eqs_v}
\end{eqnarray}
and also possess the following properties ($\overline{u} \equiv u^\dagger \gamma^0$):
\begin{equation}
\begin{aligned}
&u^\dagger_{\mathbf{p},s} u_{\mathbf{p},s'} = v^\dagger_{\mathbf{p},s} v_{\mathbf{p},s'} = \delta_{ss'} \,, \quad u^\dagger_{\mathbf{p},s} v_{\mathbf{p},s'} = 0 \,, \\
&\sum_{s=\pm 1} \big ( u_{\mathbf{p},s} u^\dagger_{\mathbf{p},s} + v_{\mathbf{p},s} v^\dagger_{\mathbf{p},s}\big ) = \mathrm{I} \,, \\
&\sum_{s=\pm 1}  u_{\mathbf{p},s} \overline{u}_{\mathbf{p},s} = \frac{\gamma^0 p^0 (\mathbf{p}) - \boldsymbol{\gamma} \mathbf{p} + m \mathrm{I}}{2p^0 (\mathbf{p})} \,, \\
&\sum_{s=\pm 1} v_{\mathbf{p},s} \overline{v}_{\mathbf{p},s} = \frac{\gamma^0 p^0 (\mathbf{p}) + \boldsymbol{\gamma} \mathbf{p} - m \mathrm{I}}{2p^0 (\mathbf{p})} \,. \\
\end{aligned}
\label{eq:uv_prop}
\end{equation}
Once $t \geqslant t_\text{in}$, the functions ${}_\zeta \chi_{\mathbf{p},s} (t)$ gain a nontrivial temporal dependence, and their evolution is governed by
\begin{equation}
\I \, {}_\zeta \dot{\chi}_{\mathbf{p},s} (t) = \big \{ \boldsymbol{\alpha} \, [ \zeta \mathbf{p} - e \mathbf{A} (t) ] + \beta m \big \} {}_\zeta \chi_{\mathbf{p},s} (t) \,.
\label{eq:chi_in_eq}
\end{equation}
The {\it out} functions ${}^\zeta \chi_{\mathbf{p},s} (t)$ satisfy the same equations. The $G$ matrices~\eqref{eq:G_inner_product} are diagonal with respect to momentum:
\begin{equation}
\begin{aligned}
G({}_+|{}^+)_{\mathbf{p},s; \, \mathbf{p}',s'} &= \delta (\mathbf{p}-\mathbf{p}') \, g({}_+|{}^+)_{\mathbf{p},s,s'} \,,\\
G({}_+|{}^-)_{\mathbf{p},s; \, \mathbf{p}',s'} &= \delta (\mathbf{p}+\mathbf{p}') \, g({}_+|{}^-)_{\mathbf{p},s,s'}\,,\\
G({}_-|{}^+)_{\mathbf{p},s; \, \mathbf{p}',s'} &= \delta (\mathbf{p}+\mathbf{p}') \, g({}_-|{}^+)_{\mathbf{p},s,s'}\,,\\
G({}_-|{}^-)_{\mathbf{p},s; \, \mathbf{p}',s'} &= \delta (\mathbf{p}-\mathbf{p}') \, g({}_-|{}^-)_{\mathbf{p},s,s'}\,,
\end{aligned}
\label{eq:G_g}
\end{equation}
where the $g$ matrices are given by
\begin{equation}
\begin{aligned}
g({}_+|{}^+)_{\mathbf{p},s,s'} &= {}_+ \chi^\dagger_{\mathbf{p},s} (t) \, {}^+ \chi_{\mathbf{p},s'} (t) \,,\\
g({}_+|{}^-)_{\mathbf{p},s,s'} &= {}_+ \chi^\dagger_{\mathbf{p},s} (t) \, {}^- \chi_{-\mathbf{p},s'} (t) \,,\\
g({}_-|{}^+)_{\mathbf{p},s,s'} &= {}_- \chi^\dagger_{\mathbf{p},s} (t) \, {}^+ \chi_{-\mathbf{p},s'} (t) \,,\\
g({}_-|{}^-)_{\mathbf{p},s,s'} &= {}_- \chi^\dagger_{\mathbf{p},s} (t) \, {}^- \chi_{\mathbf{p},s'} (t) \,.
\end{aligned}
\label{eq:g_chi}
\end{equation}
According to Eqs.~\eqref{eq:number_density_zeroth_order_G} and \eqref{eq:number_density_zeroth_order_G_pos}, the number density of particles with given momentum $\mathbf{p}$ in a spin state $s$ has the following form ($n_{\mathbf{p},s} \equiv dN_{\mathbf{p},s}/d\mathbf{p}$):
\begin{eqnarray}
\frac{(2\pi)^3}{V} n^{(e^-)}_{\mathbf{p},s} &=&  \sum_{s'} \big |g({}_-|{}^+)_{-\mathbf{p},s',s} \big |^2 \,, \label{eq:number_density_uniform_el} \\
\frac{(2\pi)^3}{V} n^{(e^+)}_{\mathbf{p},s} &=& \sum_{s'} \big |g({}_+|{}^-)_{-\mathbf{p},s',s} \big |^2 \,. \label{eq:number_density_uniform_pos}
\end{eqnarray}
Here $V$ is the volume of the system ($V \to \infty$), and we have employed a standard regularization prescription $\delta (\mathbf{p} = \mathbf{0}) = V/(2\pi)^3$, so in the case of a uniform external field, it is the number of pairs {\it per unit volume} that yields a finite value. Due to the completeness property~\eqref{eq:G_matrix_complete}, the particle densities obey
\begin{equation}
\sum_s n^{(e^-)}_{\mathbf{p},s} = \sum_s n^{(e^+)}_{-\mathbf{p},s} \,.
\label{eq:n_el_pos}
\end{equation}

In what follows, a vast part of our derivations will revolve around the concept of the {\it adiabatic} Hamiltonian eigenfunctions. Note that in the case of a uniform electric field $\mathcal{H}_e (t)$ for given $t$ is equivalent to the zero-field Hamiltonian up to a simple gauge transformation. Although the corresponding eigenfunctions can be directly found by solving the differential equation, the latter observation makes it extremely simple. We introduce the so-called {\it adiabatic basis}:
\begin{equation}
\mathcal{H}_e (t) \varphi^{(\zeta)}_{\mathbf{p},s} (\mathbf{x}; t) = \zeta \omega(\zeta \mathbf{p}, t) \varphi^{(\zeta)}_{\mathbf{p},s} (\mathbf{x}; t) \,, \quad \zeta = \pm \,,
\label{eq:adiabatic_eigenfuncions}
\end{equation}
where $\omega (\mathbf{p}, t) = p^0 (\mathbf{p} - e \mathbf{A}(t)) = \sqrt{m^2 + [\mathbf{p} - e \mathbf{A}(t)]^2}$. In Eq.~\eqref{eq:adiabatic_eigenfuncions} we use a semicolon to evidently indicate that the $t$ dependence is parametric. In the explicit form,
\begin{eqnarray}
\varphi^{(+)}_{\mathbf{p},s} (\mathbf{x}; t) &=& (2\pi)^{-3/2} \, \mathrm{e}^{\I\mathbf{p} \mathbf{x}} \, u_{\mathbf{p}-e\mathbf{A} (t),s} \,, \label{eq:adiabatic_plus}\\
\varphi^{(-)}_{\mathbf{p},s} (\mathbf{x}; t) &=& (2\pi)^{-3/2} \, \mathrm{e}^{-\I\mathbf{p} \mathbf{x}} \, v_{-\mathbf{p}-e\mathbf{A} (t),s} \,. \label{eq:adiabatic_minus}
\end{eqnarray}
The Heisenberg field operator can be decomposed according to
\begin{equation}
\psi (x) = \sum_{s} \int \! d\mathbf{p} \big [a_{\mathbf{p},s} (t) \varphi^{(+)}_{\mathbf{p},s} (\mathbf{x}; t) + b^\dagger_{\mathbf{p},s} (t) \varphi^{(-)}_{\mathbf{p},s} (\mathbf{x}; t) \big ] \,, \label{eq:psi_adiabatic}
\end{equation}
where the adiabatic creation and annihilation operators vary in time [do not confuse the adiabatic operators with those introduced within the Schr\"odinger representation in Eqs.~\eqref{eq:field_decomposition_schroedinger}]. At asymptotic times $t_\text{in/out} \to \mp \infty$, one recovers
\begin{eqnarray*}
\varphi^{(\zeta)}_{\mathbf{p},s} (\mathbf{x}; t_\text{in}) &=& {}_\zeta \varphi_{\mathbf{p},s} (\mathbf{x}) \,, \quad \varphi^{(\zeta)}_{\mathbf{p},s} (\mathbf{x}; t_\text{out}) = {}^\zeta \varphi_{\mathbf{p},s} (\mathbf{x}) \,,\\
a_{\mathbf{p},s} (t_\text{in}) &=& a_{\mathbf{p},s} (\text{in}) \,, \quad a_{\mathbf{p},s} (t_\text{out}) = a_{\mathbf{p},s} (\text{out}) \,, \\
b^\dagger_{\mathbf{p},s} (t_\text{in}) &=& b^\dagger_{\mathbf{p},s} (\text{in}) \,, \quad b^\dagger_{\mathbf{p},s} (t_\text{out}) = b^\dagger_{\mathbf{p},s} (\text{out}) \,.
\end{eqnarray*}
In the next section, we will introduce adiabatic densities of particles and deduce the system of 12 ordinary differential equations describing the pair-production process.

%%%

\begin{widetext}
\subsection{Quantum kinetic equations (QKEs)} \label{sec:qke}

The main idea is to deduce a system of equations governing the adiabatic number density and then evolve it to $t = t_\text{out}$. Although the main points of this derivation can be found in Ref.~\cite{aleksandrov_epjst_2020}, we will present it in detail in order to introduce the necessary notations in a unified manner and provide the reader with a self-contained description of the theoretical aspects.

We start with the following decomposition of the {\it in} functions in terms of the adiabatic basis taking into account the spatial homogeneity of the external field, i.e., momentum conservation:
\begin{eqnarray}
{}_- \varphi_{\mathbf{p},s} (x) &=& \sum_{s'} \Big [ A^{(1)}_{ss'} (\mathbf{p}, t) \varphi^{(+)}_{-\mathbf{p},s'} (\mathbf{x}; t) + A^{(2)}_{ss'} (\mathbf{p}, t) \varphi^{(-)}_{\mathbf{p},s'} (\mathbf{x}; t) \Big ] \,, \label{eq:in_adiabatic_minus} \\
{}_+ \varphi_{\mathbf{p},s} (x) &=& \sum_{s'} \Big [ A^{(3)}_{ss'} (\mathbf{p}, t) \varphi^{(+)}_{\mathbf{p},s'} (\mathbf{x}; t) + A^{(4)}_{ss'} (\mathbf{p}, t) \varphi^{(-)}_{-\mathbf{p},s'} (\mathbf{x}; t) \Big ] \,. \label{eq:in_adiabatic_plus}
\end{eqnarray}
By using the explicit form of the adiabatic eigenfunctions, one obtains
\begin{eqnarray}
A^{(1)}_{ss'} (\mathbf{p}, t) &=& u^\dagger_{-\mathbf{p}-e\mathbf{A} (t),s'} {}_- \chi_{\mathbf{p},s} (t) \,, \qquad A^{(2)}_{ss'} (\mathbf{p}, t) = v^\dagger_{-\mathbf{p}-e\mathbf{A} (t),s'} {}_- \chi_{\mathbf{p},s} (t) \,, \label{eq:A12_product} \\
A^{(3)}_{ss'} (\mathbf{p}, t) &=& u^\dagger_{\mathbf{p}-e\mathbf{A} (t),s'} {}_+ \chi_{\mathbf{p},s} (t) \,, \qquad A^{(4)}_{ss'} (\mathbf{p}, t) = v^\dagger_{\mathbf{p}-e\mathbf{A} (t),s'} {}_+ \chi_{\mathbf{p},s} (t) \,. \label{eq:A34_product}
\end{eqnarray}
The asymptotic values read
\begin{eqnarray}
A^{(1)}_{ss'} (\mathbf{p}, t_\text{in}) &=& 0 \,, \qquad A^{(2)}_{ss'} (\mathbf{p}, t_\text{in}) = \delta_{ss'} \,, \label{eq:A12_tin} \\
A^{(3)}_{ss'} (\mathbf{p}, t_\text{in}) &=& \delta_{ss'} \,, \qquad A^{(4)}_{ss'} (\mathbf{p}, t_\text{in}) = 0 \label{eq:A34_tin}
\end{eqnarray}
and
\begin{eqnarray}
A^{(1)}_{ss'} (\mathbf{p}, t_\text{out}) &=& g^*({}_-|{}^+)_{\mathbf{p},s,s'} \,, \qquad A^{(2)}_{ss'} (\mathbf{p}, t_\text{out}) = g^*({}_-|{}^-)_{\mathbf{p},s,s'} \,, \label{eq:A12_tout} \\
A^{(3)}_{ss'} (\mathbf{p}, t_\text{out}) &=& g^*({}_+|{}^+)_{\mathbf{p},s,s'} \,, \qquad A^{(4)}_{ss'} (\mathbf{p}, t_\text{out}) = g^*({}_+|{}^-)_{\mathbf{p},s,s'} \,. \label{eq:A34_tout}
\end{eqnarray}
According to Eqs.~\eqref{eq:number_density_uniform_el} and \eqref{eq:number_density_uniform_pos}, the particle number densities can be readily evaluated once the coefficients $A^{(1)}_{ss'}$ and $A^{(4)}_{ss'}$ are known at $t=t_\text{out}$.

By utilizing Eqs.~\eqref{eq:in_adiabatic_minus}, \eqref{eq:in_adiabatic_plus}, and the field-operator representation in terms of the {\it in} operators [general form is given in Eq.~\eqref{eq:psi_x_in}], we establish the following relations:
\begin{eqnarray}
a_{\mathbf{p},s} (t) &=& \sum_{s'} \Big [ A^{(3)}_{s's} (\mathbf{p}, t) a_{\mathbf{p},s'} (\text{in}) + A^{(1)}_{s's} (-\mathbf{p}, t) b^\dagger_{-\mathbf{p},s'} (\text{in}) \Big ] \,, \label{eq:a_adiabatic_in} \\
b^\dagger_{\mathbf{p},s} (t) &=& \sum_{s'} \Big [ A^{(4)}_{s's} (-\mathbf{p}, t) a_{-\mathbf{p},s'} (\text{in}) + A^{(2)}_{s's} (\mathbf{p}, t) b^\dagger_{\mathbf{p},s'} (\text{in}) \Big ] \,. \label{eq:b_adiabatic_in}
\end{eqnarray}

We will also need the following in-vacuum expectation values:
\begin{eqnarray}
\langle 0, \text{in} | a^\dagger_{\mathbf{p},s} (t) a_{\mathbf{p}',s'} (t) | 0,\text{in} \rangle &=& \delta (\mathbf{p} - \mathbf{p}') \sum_{s''} \big [ A^{(1)}_{s''s} (-\mathbf{p}, t) \big ]^* A^{(1)}_{s''s'} (-\mathbf{p}, t) \,, \label{eq:in_mean_ad_a} \\
\langle 0, \text{in} | b^\dagger_{\mathbf{p},s} (t) b_{\mathbf{p}',s'} (t) | 0,\text{in} \rangle &=& \delta (\mathbf{p} - \mathbf{p}') \sum_{s''} A^{(4)}_{s''s} (-\mathbf{p}, t) \big [ A^{(4)}_{s''s'} (-\mathbf{p}, t) \big ]^* \,, \label{eq:in_mean_bd_b} \\
\langle 0, \text{in} | a^\dagger_{\mathbf{p},s} (t) b^\dagger_{\mathbf{p}',s'} (t) | 0,\text{in} \rangle &=& \delta (\mathbf{p} + \mathbf{p}') \sum_{s''} \big [ A^{(1)}_{s''s} (-\mathbf{p}, t) \big ]^* A^{(2)}_{s''s'} (-\mathbf{p}, t) \,, \label{eq:in_mean_ad_bd} \\
\langle 0, \text{in} | b_{\mathbf{p},s} (t) a_{\mathbf{p}',s'} (t) | 0,\text{in} \rangle &=& \delta (\mathbf{p} + \mathbf{p}') \sum_{s''} \big [ A^{(2)}_{s''s} (\mathbf{p}, t) \big ]^* A^{(1)}_{s''s'} (\mathbf{p}, t) \,. \label{eq:in_mean_b_a}
\end{eqnarray}

Our goal is to derive a closed system of equations allowing one to evolve the functions $A^{(j)}_{ss'} (\mathbf{p}, t)$ in time. To this end, we first substitute~\eqref{eq:in_adiabatic_minus} into the Dirac equation and project the resulting equation on $\varphi^{(+)}_{-\mathbf{p},s'}$ and $\varphi^{(-)}_{\mathbf{p},s'}$:
\begin{eqnarray}
&&\I \dot{A}^{(1)}_{ss''} (\mathbf{p}, t) + \I \sum_{s'} M^{(uu)}_{s'' s'} (-\mathbf{p}, t) A^{(1)}_{ss'} (\mathbf{p}, t) \nonumber \\
&& {} + \I \sum_{s'} M^{(uv)}_{s'' s'} (-\mathbf{p}, t) A^{(2)}_{ss'} (\mathbf{p}, t) = \omega (-\mathbf{p}, t) A^{(1)}_{ss''} (\mathbf{p}, t) \,, \label{eq:system1_A1} \\
&&\I \dot{A}^{(2)}_{ss''} (\mathbf{p}, t) + \I \sum_{s'} M^{(vu)}_{s'' s'} (-\mathbf{p}, t) A^{(1)}_{ss'} (\mathbf{p}, t) \nonumber \\
&& {} + \I \sum_{s'} M^{(vv)}_{s'' s'} (-\mathbf{p}, t) A^{(2)}_{ss'} (\mathbf{p}, t) = -\omega (-\mathbf{p}, t) A^{(2)}_{ss''} (\mathbf{p}, t) \,, \label{eq:system1_A2}
\end{eqnarray}
where $M^{(uu)}_{s s'} (\mathbf{p}, t) \equiv u^\dagger_{\mathbf{p}-e\mathbf{A}(t),s} \dot{u}_{\mathbf{p}-e\mathbf{A} (t),s'}$ and the other coefficients are defined analogously. The time derivatives of the bispinors are proportional to $e\dot{\mathbf{A}} (t) \equiv e \mathbf{A}' (t) = -e \mathbf{E} (t)$. For instance,
\begin{equation}
\left. M^{(uu)}_{s s'} (\mathbf{p}, t) = e \mathbf{E} (t) (u^\dagger_{\mathbf{q},s} \boldsymbol{\nabla}_\mathbf{q} u_{\mathbf{q},s'}) \right|_{\mathbf{q} = \mathbf{p}-e\mathbf{A} (t) }\,. \label{eq:Muu_nabla}
\end{equation}
Note that $s$ in Eqs.~\eqref{eq:system1_A1} and \eqref{eq:system1_A2} is not involved in the coefficients and only governs the initial conditions for the ODE system. Let us introduce the following vector functions:
\begin{equation}
A^{(j)}_s (\mathbf{p}, t) = \begin{pmatrix}
A^{(j)}_{s,-1} (\mathbf{p}, t) \\
A^{(j)}_{s,+1} (\mathbf{p}, t)
\end{pmatrix} \,.
\end{equation}
Then the system~\eqref{eq:system1_A1}, \eqref{eq:system1_A2} can be recast into
\begin{eqnarray}
&&\I \dot{A}^{(1)}_{s} (\mathbf{p}, t) + \I M^{(uu)} (-\mathbf{p}, t) A^{(1)}_{s} (\mathbf{p}, t) \nonumber \\
&& {} + \I M^{(uv)} (-\mathbf{p}, t) A^{(2)}_{s} (\mathbf{p}, t) = \omega (-\mathbf{p}, t) A^{(1)}_{s} (\mathbf{p}, t) \,, \label{eq:system2_A1} \\
&&\I \dot{A}^{(2)}_{s} (\mathbf{p}, t) + \I M^{(vu)} (-\mathbf{p}, t) A^{(1)}_{s} (\mathbf{p}, t) \nonumber \\
&& {} + \I M^{(vv)} (-\mathbf{p}, t) A^{(2)}_{s} (\mathbf{p}, t) = -\omega (-\mathbf{p}, t) A^{(2)}_{s} (\mathbf{p}, t) \,. \label{eq:system2_A2}
\end{eqnarray}
The quantities~\eqref{eq:Muu_nabla} have become matrices with respect to the spin indices. Following Ref.~\cite{aleksandrov_epjst_2020}, we define now two-by-two matrices,
\begin{eqnarray}
\hat{A}^{(j)} (\mathbf{p}, t) &=& \sum_{s} A^{(j)}_s (\mathbf{p}, t) \big [ A^{(j)}_s (\mathbf{p}, t) \big ]^\dagger \,, \quad j = 1, \, 2 \,, \label{eq:Aj_matrices} \\
\hat{B}^{(\pm)} (\mathbf{p}, t) &=& \sum_{s} A^{(2,1)}_s (\mathbf{p}, t) \big [ A^{(1,2)}_s (\mathbf{p}, t) \big ]^\dagger \,. \label{eq:B_matrices}
\end{eqnarray}
One can straightforwardly obtain the following system (we substitute $\mathbf{p} \to - \mathbf{p}$):
\begin{eqnarray}
\dot{\hat{A}}^{(1)} (-\mathbf{p}, t) &=& - M^{(uu)} (\mathbf{p}, t) \hat{A}^{(1)} (-\mathbf{p}, t) - \hat{A}^{(1)} (-\mathbf{p}, t) \big [ M^{(uu)} (\mathbf{p}, t) \big ]^\dagger \nonumber \\
{}&-& M^{(uv)} (\mathbf{p}, t) \hat{B}^{(+)} (-\mathbf{p}, t) - \hat{B}^{(-)} (-\mathbf{p}, t) \big [ M^{(uv)} (\mathbf{p}, t) \big ]^\dagger \,, \label{eq:system3_A1} \\
\dot{\hat{A}}^{(2)} (-\mathbf{p}, t) &=& - M^{(vv)} (\mathbf{p}, t) \hat{A}^{(2)} (-\mathbf{p}, t) - \hat{A}^{(2)} (-\mathbf{p}, t) \big [ M^{(vv)} (\mathbf{p}, t) \big ]^\dagger \nonumber \\
{}&-& M^{(vu)} (\mathbf{p}, t) \hat{B}^{(-)} (-\mathbf{p}, t) - \hat{B}^{(+)} (-\mathbf{p}, t) \big [ M^{(vu)} (\mathbf{p}, t) \big ]^\dagger \,, \label{eq:system3_A2} \\
\dot{\hat{B}}^{(+)} (-\mathbf{p}, t) &=& - M^{(vv)} (\mathbf{p}, t) \hat{B}^{(+)} (-\mathbf{p}, t) - \hat{B}^{(+)} (-\mathbf{p}, t) \big [ M^{(uu)} (\mathbf{p}, t) \big ]^\dagger \nonumber \\
{}&-& M^{(vu)} (\mathbf{p}, t) \hat{A}^{(1)} (-\mathbf{p}, t) - \hat{A}^{(2)} (-\mathbf{p}, t) \big [ M^{(uv)} (\mathbf{p}, t) \big ]^\dagger + 2 \I \omega (\mathbf{p}, t) \hat{B}^{(+)} (-\mathbf{p}, t) \,, \label{eq:system3_Bplus} \\
\dot{\hat{B}}^{(-)} (-\mathbf{p}, t) &=& - M^{(uu)} (\mathbf{p}, t) \hat{B}^{(-)} (-\mathbf{p}, t) - \hat{B}^{(-)} (-\mathbf{p}, t) \big [ M^{(vv)} (\mathbf{p}, t) \big ]^\dagger \nonumber \\
{}&-& M^{(uv)} (\mathbf{p}, t) \hat{A}^{(2)} (-\mathbf{p}, t) - \hat{A}^{(1)} (-\mathbf{p}, t) \big [ M^{(vu)} (\mathbf{p}, t) \big ]^\dagger - 2 \I \omega (\mathbf{p}, t) \hat{B}^{(-)} (-\mathbf{p}, t) \,.\label{eq:system3_Bminus} 
\end{eqnarray}
Equations~\eqref{eq:system2_A1} and \eqref{eq:system2_A2} contain eight complex functions to be determined. Although the four matrices defined in Eqs.~\eqref{eq:Aj_matrices} and \eqref{eq:B_matrices} have in total 16 complex elements, the number of independent real parameters remains the same and equals 16 since $\hat{A}^{(1)}$ and $\hat{A}^{(2)}$ are Hermitian and $\big [ B^{(+)} \big ]^\dagger = \hat{B}^{(-)}$. The initial conditions are $\hat{A}^{(1)} = \hat{B}^{(\pm)} = 0$, $\hat{A}^{(2)} = \mathrm{I}$.

We note that the full information concerning the pair-production probabilities is encoded in the system~\eqref{eq:system3_A1}--\eqref{eq:system3_Bminus}, whereas substituting~\eqref{eq:in_adiabatic_plus} into the Dirac equation does not yield any additional insights. For instance, the analogous matrices~\eqref{eq:Aj_matrices} with $j=3$,~$4$ obey
\begin{eqnarray}
\hat{A}^{(3)} (\mathbf{p}, t) &=& \mathrm{I} - \hat{A}^{(1)} (-\mathbf{p}, t) \,, \label{eq:A3_A1} \\
\hat{A}^{(4)} (\mathbf{p}, t) &=& \mathrm{I} - \hat{A}^{(2)} (-\mathbf{p}, t) \,.\label{eq:A4_A2}
\end{eqnarray}
These expressions can be easily derived by utilizing the anticommutation relations in Eqs.~\eqref{eq:in_mean_ad_a} and \eqref{eq:in_mean_bd_b}. At $t = t_\text{in}$ the equations~\eqref{eq:A3_A1} and \eqref{eq:A4_A2} are obviously in agreement with the initial conditions~\eqref{eq:A12_tin} and \eqref{eq:A34_tin}, while at $t = t_\text{out}$ they correspond to the general properties~\eqref{eq:G_matrix_orthonorm} and \eqref{eq:G_matrix_complete}.

As clearly seen from Eq.~\eqref{eq:Muu_nabla}, the actual form of the $M$ matrices involved in the system~\eqref{eq:system3_A1}--\eqref{eq:system3_Bminus} depends on the choice of the bispinor basis. We will proceed as follows. First, we will employ a specific basis set and explicitly evaluate the matrices. In Sec.~\ref{sec:qke_basis}, we will discuss how unitary transformations of the bispinors affect the resulting system. Let us choose~\cite{pervushin_skokov,blaschke_prd_2011,aleksandrov_epjst_2020}
\begin{eqnarray}
u_{\boldsymbol{p},-1} = C(p^0)
\begin{pmatrix}
p^0 + m\\
0\\
p_z\\
p_x+ \I p_y
\end{pmatrix}, &\qquad& u_{\boldsymbol{p},+1} = C(p^0)
\begin{pmatrix}
0\\
p^0 + m\\
p_x - \I p_y\\
-p_z
\end{pmatrix},\label{eq:u_explicit}\\
v_{\boldsymbol{p},-1} = C(p^0)
\begin{pmatrix}
-p_z\\
-p_x - \I p_y\\
p^0 + m\\
0
\end{pmatrix}, &\qquad& v_{\boldsymbol{p},+1} = C(p^0)
\begin{pmatrix}
-p_x + \I p_y\\
p_z\\
0\\
p^0 + m
\end{pmatrix},\label{eq:v_explicit}
\end{eqnarray}
where $C(p^0) = [2p^0 (p^0 +m)]^{-1/2}$ and $p^0 = p_0 = \sqrt{m^2 + \boldsymbol{p}^2}$. These bispinors satisfy all of the relations~\eqref{eq:uv_prop}. It turns out that in this case, the necessary matrices have particularly simple form:
\begin{eqnarray}
M^{(uu)} (\mathbf{p}, t) &=& M^{(vv)} (\mathbf{p}, t) = \frac{\I e}{2q^0 (q^0+m)}\, [\mathbf{q} \times \mathbf{E} (t)] \boldsymbol{\sigma} \,, \label{eq:M_explicit_1} \\
M^{(uv)} (\mathbf{p}, t) &=& - M^{(vu)} (\mathbf{p}, t) = \frac{e}{2q_0^2 (q^0+m)}\, \big \{ [\mathbf{q} \mathbf{E}(t)] \mathbf{q} - q^0 (q^0+m) \mathbf{E} (t) \big \} \boldsymbol{\sigma} \,, \label{eq:M_explicit_2} 
\end{eqnarray}
where $\mathbf{q} = \mathbf{p}-e\mathbf{A} (t)$ is a kinetic momentum and $q^0=q_0=\sqrt{m^2 + \mathbf{q}^2} = \omega(\mathbf{p}, t)$. The matrices in the first (second) line are anti-Hermitian (Hermitian). This property is basis-independent. Let us define the vector functions $\boldsymbol{\mu}_1$ and $\boldsymbol{\mu}_2$ via
\begin{equation}
M^{(uu)} (\mathbf{p}, t) = \I \boldsymbol{\mu}_1 (\mathbf{p}, t) \boldsymbol{\sigma} \,, \quad M^{(uv)} (\mathbf{p}, t) = \boldsymbol{\mu}_2 (\mathbf{p}, t) \boldsymbol{\sigma} \,. \label{eq:mu_def}
\end{equation}
Using the explicit form of the matrices, one rewrites Eqs.~\eqref{eq:system3_A1}--\eqref{eq:system3_Bminus} as follows:
\begin{eqnarray}
\dot{\hat{A}}^{(1)} (-\mathbf{p}, t) &=& \I \boldsymbol{\mu}_1 (\mathbf{p}, t) \big [ \hat{A}^{(1)} (-\mathbf{p}, t), \, \boldsymbol{\sigma} \big ] \nonumber \\
{}&-& \boldsymbol{\mu}_2 (\mathbf{p}, t) \big [ \boldsymbol{\sigma} \hat{B}^{(+)} (-\mathbf{p}, t) + \hat{B}^{(-)} (-\mathbf{p}, t) \boldsymbol{\sigma} \big ] \,, \label{eq:system4_A1} \\
\dot{\hat{A}}^{(2)} (-\mathbf{p}, t) &=& \I \boldsymbol{\mu}_1 (\mathbf{p}, t) \big [ \hat{A}^{(2)} (-\mathbf{p}, t), \, \boldsymbol{\sigma} \big ] \nonumber \\
{}&+& \boldsymbol{\mu}_2 (\mathbf{p}, t) \big [ \boldsymbol{\sigma} \hat{B}^{(-)} (-\mathbf{p}, t) + \hat{B}^{(+)} (-\mathbf{p}, t) \boldsymbol{\sigma} \big ] \,, \label{eq:system4_A2} \\
\dot{\hat{B}}^{(+)} (-\mathbf{p}, t) &=&\I \boldsymbol{\mu}_1 (\mathbf{p}, t) \big [ \hat{B}^{(+)} (-\mathbf{p}, t), \, \boldsymbol{\sigma} \big ] \nonumber \\
{}&+& \boldsymbol{\mu}_2 (\mathbf{p}, t) \big [ \boldsymbol{\sigma} \hat{A}^{(1)} (-\mathbf{p}, t) - \hat{A}^{(2)} (-\mathbf{p}, t) \boldsymbol{\sigma} \big ] + 2 \I \omega (\mathbf{p}, t) \hat{B}^{(+)} (-\mathbf{p}, t) \,, \label{eq:system4_Bplus} \\
\dot{\hat{B}}^{(-)} (-\mathbf{p}, t) &=& \I \boldsymbol{\mu}_1 (\mathbf{p}, t) \big [ \hat{B}^{(-)} (-\mathbf{p}, t), \, \boldsymbol{\sigma} \big ] \nonumber \\
{}&-& \boldsymbol{\mu}_2 (\mathbf{p}, t) \big [ \boldsymbol{\sigma} \hat{A}^{(2)} (-\mathbf{p}, t) - \hat{A}^{(1)} (-\mathbf{p}, t) \boldsymbol{\sigma} \big ]  - 2 \I \omega (\mathbf{p}, t) \hat{B}^{(-)} (-\mathbf{p}, t) \,.\label{eq:system4_Bminus} 
\end{eqnarray}
It is then natural and convenient to expand the matrices in terms of the Pauli ones, but let us first define the following Hermitian combinations of non-Hermitian matrices $\hat{B}^{(\pm)}$:
\begin{eqnarray}
\hat{U} (\mathbf{p}, t) &=& \frac{1}{2} \, \big [ \hat{B}^{(+)} (\mathbf{p}, t) + \hat{B}^{(-)} (\mathbf{p}, t) \big ] \,, \label{eq:U_matrix} \\
\hat{V} (\mathbf{p}, t) &=& \frac{\I}{2} \, \big [ \hat{B}^{(+)} (\mathbf{p}, t) - \hat{B}^{(-)} (\mathbf{p}, t) \big ] \,. \label{eq:V_matrix} 
\end{eqnarray}
Then we introduce
\begin{equation}
\hat{A}^{(1)} (-\mathbf{p}, t) = a^{(1)} (\mathbf{p}, t) \mathrm{I} + \mathbf{a}^{(1)} (\mathbf{p}, t) \boldsymbol{\sigma} \,.
\label{eq:Pauli_A1}
\end{equation}
Similar notations will be used also for $\hat{A}^{(2)}$,  $\hat{U}$, and $\hat{V}$. We always assume that the elements $\boldsymbol{
\sigma}_{ss'}$ of the Pauli matrices are indexed by $s$, $s' = \pm 1$. The coefficients are real and can be evaluated via
\begin{equation}
a^{(1)} (\mathbf{p}, t) = \frac{1}{2} \, \mathrm{Tr} \, \hat{A}^{(1)} (-\mathbf{p}, t) \,, \quad  \mathbf{a}^{(1)} (\mathbf{p}, t) = \frac{1}{2} \, \mathrm{Tr} \, \big [ \hat{A}^{(1)} (-\mathbf{p}, t) \boldsymbol{\sigma} \big ] \,.
\label{eq:Pauli_coeff}
\end{equation}
The system~\eqref{eq:system4_A1}--\eqref{eq:system4_Bminus} takes now the form
\begin{eqnarray}
\dot{a}^{(1)} &=& -2 \boldsymbol{\mu}_2 \mathbf{u} \,, \label{eq:system_final_a1_s}\\
\dot{a}^{(2)} &=& 2 \boldsymbol{\mu}_2 \mathbf{u} \,, \label{eq:system_final_a2_s} \\
\dot{u} &=& \boldsymbol{\mu}_2 (\mathbf{a}^{(1)} - \mathbf{a}^{(2)}) + 2 \omega v \,, \label{eq:system_final_u_s} \\
\dot{v} &=& - 2 \omega u \,, \label{eq:system_final_v_s} \\
\dot{\mathbf{a}}^{(1)} &=& 2 (\boldsymbol{\mu}_1 \times \mathbf{a}^{(1)}) - 2 (\boldsymbol{\mu}_2 \times \mathbf{v}) - 2 \boldsymbol{\mu}_2 u \,, \label{eq:system_final_a1_v} \\
\dot{\mathbf{a}}^{(2)} &=& 2 (\boldsymbol{\mu}_1 \times \mathbf{a}^{(2)}) - 2 (\boldsymbol{\mu}_2 \times \mathbf{v}) + 2 \boldsymbol{\mu}_2 u \,, \label{eq:system_final_a2_v} \\
\dot{\mathbf{u}} &=& 2 (\boldsymbol{\mu}_1 \times \mathbf{u}) + \boldsymbol{\mu}_2 (a^{(1)} - a^{(2)}) + 2 \omega \mathbf{v} \,, \label{eq:system_final_u_v} \\
\dot{\mathbf{v}} &=& 2 (\boldsymbol{\mu}_1 \times \mathbf{v}) - \boldsymbol{\mu}_2 \times (\mathbf{a}^{(1)} + \mathbf{a}^{(2)}) - 2 \omega \mathbf{u} \,. \label{eq:system_final_v_v}
\end{eqnarray}
Here all of the functions have arguments $\mathbf{p}$ and $t$. The initial conditions yield $a^{(2)} ( \mathbf{p}, t_\text{in}) = 1$, while the other functions vanish.

The quantities~\eqref{eq:in_mean_ad_a}--\eqref{eq:in_mean_b_a} in terms of the QKE components read
\begin{eqnarray}
\begin{aligned}
\langle 0, \text{in} | a^\dagger_{\mathbf{p},s} (t) a_{\mathbf{p}',s'} (t) | 0,\text{in} \rangle &= \delta (\mathbf{p} - \mathbf{p}') \big [ a^{(1)} (\mathbf{p}, t) \delta_{s's} + \mathbf{a}^{(1)} (\mathbf{p}, t) \boldsymbol{\sigma}_{s's} \big ] \,, \\
\langle 0, \text{in} | b^\dagger_{\mathbf{p},s} (t) b_{\mathbf{p}',s'} (t) | 0,\text{in} \rangle &= \delta (\mathbf{p} - \mathbf{p}') \big \{ [1- a^{(2)} (-\mathbf{p}, t)] \delta_{ss'} - \mathbf{a}^{(2)} (-\mathbf{p}, t) \boldsymbol{\sigma}_{ss'} \big \} \,, \\
\langle 0, \text{in} | a^\dagger_{\mathbf{p},s} (t) b^\dagger_{\mathbf{p}',s'} (t) | 0,\text{in} \rangle &= \delta (\mathbf{p} + \mathbf{p}') \big \{ [ u (\mathbf{p}, t) - \I v (\mathbf{p}, t)] \delta_{s's} + [\mathbf{u} (\mathbf{p}, t) - \I \mathbf{v} (\mathbf{p}, t) ] \boldsymbol{\sigma}_{s's} \big \} \,, \\
\langle 0, \text{in} | b_{\mathbf{p},s} (t) a_{\mathbf{p}',s'} (t) | 0,\text{in} \rangle &= \delta (\mathbf{p} + \mathbf{p}') \big \{ [ u (\mathbf{p}', t) + \I v (\mathbf{p}', t)] \delta_{s's} + [\mathbf{u} (\mathbf{p}', t) + \I \mathbf{v} (\mathbf{p}', t) ] \boldsymbol{\sigma}_{s's} \big \} \,.
\end{aligned}
\label{eq:vac_mean_1}
\end{eqnarray}
The number densities~\eqref{eq:number_density_uniform_el}, \eqref{eq:number_density_uniform_pos} of the particles produced can be then calculated according to
\begin{eqnarray}
\begin{aligned}
\frac{(2\pi)^3}{V} n^{(e^-)}_{\mathbf{p},s} &= a^{(1)} (\mathbf{p}, t_\text{out}) - (\mathrm{sign} \, s) \, a^{(1)}_3 (\mathbf{p}, t_\text{out}) \,, \\
\frac{(2\pi)^3}{V} n^{(e^+)}_{\mathbf{p},s} &= 1 - a^{(2)} (-\mathbf{p}, t_\text{out}) + (\mathrm{sign} \, s) \, a^{(2)}_3 (-\mathbf{p}, t_\text{out}) \,.
\end{aligned}
\label{eq:number_density_qke_1}
\end{eqnarray}
The equations contain only the third components of $\mathbf{a}^{(1)}$ and $\mathbf{a}^{(2)}$ since only the Pauli matrix $\sigma_z$ has nonzero diagonal elements. It turns out that the second term in the right-hand side of Eq.~\eqref{eq:system_final_a2_v} is opposite to the analogous contribution presented in Ref.~\cite{pervushin_skokov} [second term in the right-hand side of Eq.~(46)]. In Ref.~\cite{pervushin_skokov} the sum of Eqs.~\eqref{eq:system_final_a1_v} and \eqref{eq:system_final_a2_v} yielded a homogeneous equation leading to $\mathbf{a}^{(1)} = -\mathbf{a}^{(2)}$. Together with an obvious relation $a^{(2)} = 1 - a^{(1)}$ following from Eqs.~\eqref{eq:system_final_a1_s} and \eqref{eq:system_final_a2_s}, the authors of Ref.~\cite{pervushin_skokov} came to a system of 12 equations involving the functions $a^{(1)}$, $u$, $v$, $\mathbf{a}^{(1)}$, $\mathbf{u}$, and $\mathbf{v}$ (in our notation). This system of quantum kinetic equations (QKEs) has been extensively discussed in the literature (see, e.g., Refs.~\cite{pervushin_skokov, aleksandrov_epjst_2020} and references therein). It appears that the derivation given in the original paper~[46] up to now was never {\it fully} confirmed [for instance, Refs.~\cite{blaschke_prd_2011,aleksandrov_epjst_2020} contain a correct and detailed derivation of the intermediate system, which is absolutely compatible with our Eqs.~\eqref{eq:system4_A1}--\eqref{eq:system4_Bminus}]. Now it is clear that the correct version of the QKEs is different as the sum of Eqs.~\eqref{eq:system_final_a1_v} and \eqref{eq:system_final_a2_v} is, in fact, nonhomogeneous. Although upon first glance, it may be unclear whether one can significantly reduce the number of the unknown functions in Eqs.~\eqref{eq:system_final_a1_s}--\eqref{eq:system_final_v_v}, the system can be simplified even more greatly. One can easily verify that $u$, $v$, and the difference $\mathbf{a}^{(1)} - \mathbf{a}^{(2)}$ form a closed homogeneous subsystem, so they vanish.

By substituting now $f \equiv a^{(1)}$, $a^{(2)} = 1 - f$, $\mathbf{f} \equiv \mathbf{a}^{(1)} = \mathbf{a}^{(2)}$, and $u = v = 0$, we obtain the following QKEs involving ten unknown components:
\begin{eqnarray}
\dot{f} &=& -2 \boldsymbol{\mu}_2 \mathbf{u} \,, \label{eq:system_simple_f_s}\\
\dot{\mathbf{f}} &=& 2 (\boldsymbol{\mu}_1 \times \mathbf{f}) - 2 (\boldsymbol{\mu}_2 \times \mathbf{v}) \,, \label{eq:system_simple_f_v} \\
\dot{\mathbf{u}} &=& 2 (\boldsymbol{\mu}_1 \times \mathbf{u}) + \boldsymbol{\mu}_2 (2f - 1) + 2 \omega \mathbf{v} \,, \label{eq:system_simple_u_v} \\
\dot{\mathbf{v}} &=& 2 (\boldsymbol{\mu}_1 \times \mathbf{v}) - 2 (\boldsymbol{\mu}_2 \times \mathbf{f}) - 2 \omega \mathbf{u} \,. \label{eq:system_simple_v_v}
\end{eqnarray}
All of the functions vanish at $t = t_\text{in}$. The system is manifestly gauge invariant (one should only keep in mind the difference between the kinetic and generalized momentum). The QKEs possess the following integral of motion:
\begin{equation}
\frac{1}{4} (1-2f)^2 + \mathbf{f}^2 + \mathbf{u}^2 + \mathbf{v}^2 = \frac{1}{4} \,.
\end{equation}
From this it follows that the function $f$ never exceeds unity in accordance with the Pauli exclusion principle. The relations~\eqref{eq:vac_mean_1} now read
\begin{eqnarray}
\begin{aligned}
\langle 0, \text{in} | a^\dagger_{\mathbf{p},s} (t) a_{\mathbf{p}',s'} (t) | 0,\text{in} \rangle &= \delta (\mathbf{p} - \mathbf{p}') \big [ f (\mathbf{p}, t) \delta_{s's} + \mathbf{f} (\mathbf{p}, t) \boldsymbol{\sigma}_{s's} \big ] \,, \\
\langle 0, \text{in} | b^\dagger_{\mathbf{p},s} (t) b_{\mathbf{p}',s'} (t) | 0,\text{in} \rangle &= \delta (\mathbf{p} - \mathbf{p}') \big [ f (-\mathbf{p}, t) \delta_{ss'} - \mathbf{f} (-\mathbf{p}, t) \boldsymbol{\sigma}_{ss'} \big ] \,, \\
\langle 0, \text{in} | a^\dagger_{\mathbf{p},s} (t) b^\dagger_{\mathbf{p}',s'} (t) | 0,\text{in} \rangle &= \delta (\mathbf{p} + \mathbf{p}') [\mathbf{u} (\mathbf{p}, t) - \I \mathbf{v} (\mathbf{p}, t) ] \boldsymbol{\sigma}_{s's} \,, \\
\langle 0, \text{in} | b_{\mathbf{p},s} (t) a_{\mathbf{p}',s'} (t) | 0,\text{in} \rangle &= \delta (\mathbf{p} + \mathbf{p}') [\mathbf{u} (-\mathbf{p}, t) + \I \mathbf{v} (-\mathbf{p}, t) ] \boldsymbol{\sigma}_{s's} \,.
\end{aligned}
\label{eq:vac_mean_2}
\end{eqnarray}
The spin-resolved number densities are given by
\begin{eqnarray}
\frac{(2\pi)^3}{V} n^{(e^-)}_{\mathbf{p},s} &=& f (\mathbf{p}, t_\text{out}) - (\mathrm{sign} \, s) \, f_z (\mathbf{p}, t_\text{out}) \,, \label{eq:number_density_qke_el} \\
\frac{(2\pi)^3}{V} n^{(e^+)}_{\mathbf{p},s} &=& f (-\mathbf{p}, t_\text{out}) + (\mathrm{sign} \, s) \, f_z (-\mathbf{p}, t_\text{out}) \,. \label{eq:number_density_qke_pos}
\end{eqnarray}
Note that the kinetic formalism yields the distribution functions which are completely compatible with Eq.~\eqref{eq:n_el_pos}. We also point out that the spin anisotropy governed by the third component of the vector function $\mathbf{f}$ is determined here with respect to the bispinor basis~\eqref{eq:u_explicit}. A more detailed discussion of the spin effects is presented in Sec.~\ref{sec:helicity}.

It was believed that the QKE system contains a closed subsystem involving $f$, $\mathbf{u}$, and $\mathbf{v}$, which coincides with Eqs.~\eqref{eq:system_simple_f_s}--\eqref{eq:system_simple_v_v} once one sets $\mathbf{f} = \mathbf{0}$ (see, e.g., Refs.~\cite{pervushin_skokov, aleksandrov_epjst_2020}). We see now that the correct equations are, in fact, coupled to each other and the solutions $f$, $\mathbf{u}$, and $\mathbf{v}$ are different as soon as $\mathbf{f} \neq \mathbf{0}$. Next, we will consider the reduction of the QKEs in the case of a linearly polarized external field and show that $\mathbf{f} = \mathbf{0}$ if and only if the vector $\mathbf{E} (t)$ does not rotate. In this simpler case, we will recover a well-known three-component system~\cite{GMM,pervushin_skokov,schmidt_1998,kluger_prd_1998}.

\end{widetext}

%%%

\subsection{Linearly polarized electric field} \label{sec:qke_LP}

Let us introduce a unit vector $\mathbf{n} (t)$ via $\mathbf{E} (t) = E(t) \mathbf{n} (t)$. Here we will demonstrate that once $\mathbf{n} = \mathrm{const}$, the system~\eqref{eq:system_simple_f_s}--\eqref{eq:system_simple_v_v} can be formulated in terms of three real-valued functions. To this end, we will follow the projection method proposed in Ref.~\cite{aleksandrov_epjst_2020} and applied to the previously-used (incorrect) QKEs. The main idea is to introduce a unit vector $\mathbf{e} (t)$ according to $\boldsymbol{\mu}_2 (t) = \mu_2 (t) \mathbf{e} (t)$ and to decompose $\mathbf{f}$, $\mathbf{u}$, and $\mathbf{v}$ in terms of the longitudinal and transverse components with respect to $\mathbf{e} (t)$: $\mathbf{f} = f_\parallel \mathbf{e} + \mathbf{f}_\perp$, where $f_\parallel = \mathbf{f} \mathbf{e}$ (similar notations are used for $\mathbf{u}$ and $\mathbf{v}$).

In Appendix~\ref{app:LP}, we prove the following statement:
\begin{equation}
\dot{\mathbf{n}} = \mathbf{0} \quad \Longleftrightarrow \quad \mathbf{f} = \mathbf{0}~~\text{and}~~\mathbf{u}_\perp = \mathbf{v}_\perp = \mathbf{0} \,.
\label{eq:LP_statement}
\end{equation}
In the case of a linearly polarized electric field, the QKEs are given by (see Appendix~\ref{app:LP})
\begin{eqnarray}
\begin{aligned}
\dot{f} &= -2 \mu_2 u_\parallel \,, \\
\dot{u}_\parallel &= \mu_2 (2f - 1) + 2\omega v_\parallel \,, \\
\dot{v}_\parallel &= -2\omega u_\parallel \,,
\end{aligned}
\label{eq:qke_LP}
\end{eqnarray}
where
\begin{equation}
\mu_2 = \mu_2 (\mathbf{p}, t) = \frac{eE(t) \sqrt{q_0^2 - (\mathbf{q} \mathbf{n})^2}}{2q_0^2} =  \frac{eE(t) \pi_\perp}{2\omega^2 (\mathbf{p} , t)} \,.
\end{equation}
Here $\pi_\perp \equiv \sqrt{m^2 + \mathbf{p}_\perp^2}$, and $\mathbf{p}_\perp = \mathbf{q}_\perp$ is the transverse momentum component, i.e. that perpendicular to $\mathbf{n}$. The system~\eqref{eq:qke_LP} has been discussed and implemented in numerous studies (see, e.g., Refs.~\cite{sevostyanov_prd_2021, schmidt_1998, kluger_prd_1998, schmidt_prd_1999, bloch_prd_1999, blaschke_prd_2011, aleksandrov_epjst_2020, alkofer_prl_2001, otto_plb_2015, panferov_epjd_2016, aleksandrov_symmetry, aleksandrov_sevostyanov_2022}). It is often formulated in terms of $\tilde{f} = f$, $\tilde{u} = -2u_\parallel$, and $\tilde{v} = 2 v_\parallel$. This system was also obtained from the previously-used (incorrect) general QKEs (see, e.g., Ref.~\cite{aleksandrov_epjst_2020}).

The system~\eqref{eq:system_simple_f_s}--\eqref{eq:system_simple_v_v} originates from a specific choice of the bispinor basis~\eqref{eq:u_explicit}--\eqref{eq:v_explicit} leading to a particularly simple form of the matrices~\eqref{eq:M_explicit_1}--\eqref{eq:M_explicit_2}. Next, we will discuss the alternative forms of the QKEs which appear when one employs different sets of the bispinors.

%%%

\subsection{Bispinor basis and the ambiguity of the QKEs} \label{sec:qke_basis}

Let us consider a linear transformation of the bispinors~\eqref{eq:u_explicit}--\eqref{eq:v_explicit}:
\begin{equation}
u_{\mathbf{p},s} = \sum_{s' = \pm} \alpha_{s's} \tilde{u}_{\mathbf{p},s'} \,, \qquad v_{\mathbf{p},s} = \sum_{s' = \pm} \beta_{s's} \tilde{v}_{\mathbf{p},s'} \,.
\end{equation}
In order for the tilded bispinors to satisfy the conditions of orthonormality and completeness [see Eqs.~\eqref{eq:uv_prop}], the two-by-two matrices $\alpha$ and $\beta$ should be unitary, $\alpha^\dagger \alpha = \alpha \alpha^\dagger = \mathrm{I}$. We can also choose the overall phase, so that the determinants of $\alpha$ and $\beta$ equal unity, i.e. $\alpha, \, \beta \in \mathrm{SU} (2)$. The $M$ matrices in Eqs.~\eqref{eq:M_explicit_1}--\eqref{eq:M_explicit_2} are then transformed according to
\begin{eqnarray}
\begin{aligned}
\tilde{M}^{(uu)} &= \alpha M^{(uu)} \alpha^\dagger \,, \qquad \tilde{M}^{(uv)} = \alpha M^{(uv)} \beta^\dagger \,, \\
\tilde{M}^{(vu)} &= \beta M^{(vu)} \alpha^\dagger \,, \qquad \tilde{M}^{(vv)} = \beta M^{(vv)} \beta^\dagger \,.
\end{aligned}
\label{eq:M_tilde}
\end{eqnarray}
It is convenient to use the following representation:
\begin{equation}
\alpha = \mathfrak{a}\mathrm{I} + \I (\boldsymbol{\mathfrak{a}} \boldsymbol{\sigma}) \,, \qquad \beta = \mathfrak{b} \mathrm{I} + \I (\boldsymbol{\mathfrak{b}} \boldsymbol{\sigma}) \,, 
\end{equation}
where the coefficients are real and obey $\mathfrak{a}^2 + \boldsymbol{\mathfrak{a}}^2 = \mathfrak{b}^2 + \boldsymbol{\mathfrak{b}}^2 = 1$. One can straightforwardly show that
\begin{eqnarray}
\alpha \boldsymbol{\sigma} \beta^\dagger &=& [ \mathfrak{a} \mathfrak{b} -(\boldsymbol{\mathfrak{a}}\boldsymbol{\mathfrak{b}})] \boldsymbol{\sigma} - \I (\mathfrak{a} \boldsymbol{\mathfrak{b}} - \mathfrak{b} \boldsymbol{\mathfrak{a}}) \mathrm{I} + \boldsymbol{\mathfrak{a}} (\boldsymbol{\mathfrak{b}} \boldsymbol{\sigma}) + \boldsymbol{\mathfrak{b}} (\boldsymbol{\mathfrak{a}} \boldsymbol{\sigma}) \nonumber \\
{}&+& (\mathfrak{a} \boldsymbol{\mathfrak{b}} + \mathfrak{b} \boldsymbol{\mathfrak{a}}) \times \boldsymbol{\sigma} - \I (\boldsymbol{\mathfrak{a}} \times \boldsymbol{\mathfrak{b}})\mathrm{I}\,.
\end{eqnarray}
From this expression, one can readily obtain also $\alpha \boldsymbol{\sigma} \alpha^\dagger$, $\beta \boldsymbol{\sigma} \beta^\dagger$, and $\beta \boldsymbol{\sigma} \alpha^\dagger$.

We note that for transformations corresponding to arbitrary $\mathfrak{a}$, $\mathfrak{b}$, $\boldsymbol{\mathfrak{a}}$, and $\boldsymbol{\mathfrak{b}}$, the QKE system has a very complicated form, which can hardly simplify the computations. If $\boldsymbol{\mathfrak{a}} = \boldsymbol{\mathfrak{b}} = \mathbf{0}$, the unitary transformation can only change the signs of the matrices, so it is trivial. However, one may ask whether $\mathfrak{a} = \mathfrak{b} = 0$, $\boldsymbol{\mathfrak{a}} = \boldsymbol{\mathfrak{b}}$ can be helpful. In this case, for instance, Eqs.~\eqref{eq:system_final_a1_s} and \eqref{eq:system_final_a1_v} take the form
\begin{eqnarray}
\dot{a}^{(1)} &=& 2 \boldsymbol{\mu}_2 \mathbf{u} - 4 (\boldsymbol{\mu}_2 \boldsymbol{\mathfrak{a}}) (\boldsymbol{\mathfrak{a}} \mathbf{u}) \,, \label{eq:rot_a1_s}\\
\dot{\mathbf{a}}^{(1)} &=& - 2 (\boldsymbol{\mu}_1 \times \mathbf{a}^{(1)}) + 2\boldsymbol{\mu}_2 u + 4 (\boldsymbol{\mu}_1 \boldsymbol{\mathfrak{a}})(\boldsymbol{\mathfrak{a}} \times \mathbf{a}^{(1)}) \nonumber \\
{}&-& 4 (\boldsymbol{\mu}_2 \boldsymbol{\mathfrak{a}}) \boldsymbol{\mathfrak{a}} u - 4(\boldsymbol{\mu}_2 \boldsymbol{\mathfrak{a}})(\boldsymbol{\mathfrak{a}} \times \mathbf{v}) + 2 (\boldsymbol{\mu}_2 \times \mathbf{v}) \,.
\label{eq:rot_a1_v}
\end{eqnarray}
Since the unitary transformation does not depend on time, the vector $\boldsymbol{\mathfrak{a}}$ is constant, so its direction should be governed by a constant vector, which for arbitrary field polarization can only be that of $\mathbf{p}$. Nevertheless, from Eqs.~\eqref{eq:rot_a1_s} and \eqref{eq:rot_a1_v} one already sees that the alternative QKE system turns out to be more complicated than~\eqref{eq:system_simple_f_s}--\eqref{eq:system_simple_v_v}. Although the QKE indeed depend on the choice of the bispinor basis, it is more convenient to opt for the system~\eqref{eq:system_simple_f_s}--\eqref{eq:system_simple_v_v}.

%%%%%%%%%%%%%%%%%%%%%%%%%%%%%%%%%%%%%%%%%%%%%%%%%%%%%%%%%

\section{Dirac-Heisenberg-Wigner (DHW) formalism} \label{sec:dhw}

In this section, we will scrutinize the DHW formalism~\cite{vasak_1987,BB_prd_1991,zhuang_1996,zhuang_prd_1998,ochs_1998,hebenstreit_prd_2010} and evidently demonstrate how this approach is related to the QKE and Furry-picture quantization. First, we will introduce the concept of the Wigner function and briefly recall how one can describe electron-positron pair production in an arbitrary space-time-dependent external field. The remaining part of this section will contain our main findings concerning kinetic theory in uniform electric fields. In particular, we will show that the DHW formalism is equivalent to the ten-dimensional QKE system~\eqref{eq:system_simple_f_s}--\eqref{eq:system_simple_v_v} and discuss the spin effects.

%%%

\subsection{General approach in four dimensions} \label{sec:dhw_gen}

Although the Wigner function can be used to develop kinetic theory in a fully covariant form, according to the setup employed throughout this paper, we assume that the external electric field is specified in a certain reference frame together with the time instants $t_\text{in}$ and $t_\text{out}$. This approach corresponds to the so-called equal-time DHW formalism leading to an initial-value problem~\cite{BB_prd_1991,zhuang_1996,zhuang_prd_1998,ochs_1998,hebenstreit_prd_2010}.

Assuming that the external electromagnetic field is described in the temporal gauge $A_0 = 0$, let us define the following operator:
\begin{eqnarray}
\hat{C} (\mathbf{x}, \mathbf{s}, t) &=& \mathrm{exp} \bigg [ - \I e \int_{-1/2}^{1/2} \mathbf{A} (t, \mathbf{x} + \lambda \mathbf{s}) \mathbf{s} \, d\lambda \bigg ] \nonumber \\
{}&\times &\big [ \psi(t, \mathbf{x} + \mathbf{s}/2), \, \overline{\psi}(t, \mathbf{x} - \mathbf{s}/2) \big ] \,,
\label{eq:C_def}
\end{eqnarray}
where $\psi (t, \mathbf{x})$ is the Heisenberg field operator defined in Eq.~\eqref{eq:psi_heis}, and the commutator involves only the creation and annihilation operators. The Fourier transform with respect to $\mathbf{s}$ is called the Wigner operator,
\begin{equation}
\hat{W} (\mathbf{x}, \mathbf{p}, t) = - \frac{1}{2} \int \hat{C} (\mathbf{x}, \mathbf{s}, t) \, \mathrm{e}^{-\I \mathbf{p} \mathbf{s}} d \mathbf{s} \,.
\label{eq:W_op_s}
\end{equation}
In usual one-particle quantum mechanics, the analogous notations are introduced in terms of the wave function or density operator (see, e.g., Ref.~\cite{case_2008} and references therein). We also underline that we neglect the quantized part of the electromagnetic field, so the electron-positron field $\psi$ interacts only with the external background $\mathbf{A}$. The Wilson line factor, i.e., the exponential function in Eq.~\eqref{eq:C_def}, is usually introduced in order to describe the particle spectra in terms of the kinetic (gauge-invariant) momentum as will be seen below.

The matrix-valued Wigner function is then defined via the {\it in}-vacuum expectation value,
\begin{equation}
W (\mathbf{x}, \mathbf{p}, t) = \langle 0,\text{in} | \hat{W} (\mathbf{x}, \mathbf{p}, t) |0,\text{in} \rangle  \,.
\label{eq:wigner_function_def}
\end{equation}
Next, let us decompose the Wigner function in terms of the basis of the Clifford algebra:   
\begin{equation*}
W = \frac{1}{4} \Big [ \mathfrak{s} \mathrm{I} + \I \mathfrak{p} \gamma^5 + \mathfrak{v}_\mu \gamma^\mu + \mathfrak{a}_\mu \gamma^\mu \gamma^5 + \mathfrak{t}^{(1)}_i \sigma^{0i} + \frac{1}{2} \varepsilon^{ijk} \mathfrak{t}^{(2)}_k \sigma_{ij} \Big ] \,.
\end{equation*}
Here $\gamma^5 = \I \gamma^0 \gamma^1 \gamma^2 \gamma^3$, $\sigma^{\mu\nu} = (\I/2) [\gamma^\mu, \, \gamma^\nu]$, and the coefficients also depend on $\mathbf{x}$, $\mathbf{p}$, and $t$. They can be found via
\begin{equation}
\begin{aligned}
\mathfrak{s} &= \mathrm{Tr} \, W \,, \quad \mathfrak{p} = -\I \mathrm{Tr} \, (\gamma^5 W) \,, \\
\mathfrak{v}^\mu &=  \mathrm{Tr} \, (\gamma^\mu W) \,, \quad \mathfrak{a}^\mu =  \mathrm{Tr} \, (\gamma^5 \gamma^\mu W) \,,\\
\mathfrak{t}^{(1)}_i &= - \mathrm{Tr} \, (\sigma^{0i} W) \,, \quad \mathfrak{t}^{(2)}_i = \frac{1}{2} \varepsilon_{ijk} \mathrm{Tr} \, (\sigma^{jk} W) \,.
\end{aligned}
\label{eq:wigner_coeff}
\end{equation}
Although the Wigner function $W$ is not Hermitian, the coefficients~\eqref{eq:wigner_coeff} are real since $W^\dagger = \gamma^0 W \gamma^0$ similarly to the property of $\gamma^\mu$. In the zero-field limit, one can easily evaluate Eq.~\eqref{eq:wigner_function_def} by means of the expansion~\eqref{eq:psi_x_in}, where the {\it in} solutions coincide with the {\it out} ones and simply represent the solutions of the free Dirac equation. In this case, one obtains
\begin{equation}
W_{\mathbf{A}=\mathbf{0}} (\mathbf{p}) = -\frac{1}{2} \sum_{s} \big [ u_{\mathbf{p},s} \overline{u}_{\mathbf{p},s} - v_{\mathbf{p},s} \overline{v}_{\mathbf{p},s} \big ] = \frac{\boldsymbol{\gamma} \mathbf{p} - m\mathrm{I}}{2p^0 (\mathbf{p})} \,,
\label{eq:wigner_zero_field_W}
\end{equation}
where we have used Eqs.~\eqref{eq:uv_prop}. This means that in the absence of the external field,
\begin{equation}
\mathfrak{s}_{\mathbf{A}=\mathbf{0}} (\mathbf{p}) = -\frac{2m}{p^0 (\mathbf{p})} \,, \qquad \boldsymbol{\mathfrak{v}}_{\mathbf{A}=\mathbf{0}} (\mathbf{p}) = - \frac{2\mathbf{p}}{p^0 (\mathbf{p})} \,,
\label{eq:wigner_zero_field_sv}
\end{equation}
where $\boldsymbol{\mathfrak{v}}$ is the spatial part of $\mathfrak{v}^\mu = (\mathfrak{v}^0, \, \boldsymbol{\mathfrak{v}})$, and the other components vanish. When the external field is switched on, the temporal evolution of $W (\mathbf{x}, \mathbf{p}, t)$ is governed by the system of integro-differential equations which can be derived from the Dirac equation for $\psi (x)$. It has the following form~\cite{BB_prd_1991}:
\begin{eqnarray}
D_t \mathfrak{s} &=& 2 \boldsymbol{\Pi} \boldsymbol{\mathfrak{t}}^{(1)} \,, \label{eq:dhw_s} \\
D_t \mathfrak{p} &=& -2 \boldsymbol{\Pi} \boldsymbol{\mathfrak{t}}^{(2)} -2m\mathfrak{a}^0\,, \label{eq:dhw_p} \\
D_t \mathfrak{v}^0 &=& - \boldsymbol{D} \boldsymbol{\mathfrak{v}} \,, \label{eq:dhw_v0} \\
D_t \mathfrak{a}^0 &=& - \boldsymbol{D} \boldsymbol{\mathfrak{a}} + 2 m \mathfrak{p} \,, \label{eq:dhw_a0} \\
D_t \boldsymbol{\mathfrak{v}} &=& - \boldsymbol{D} \mathfrak{v}^0 - 2 \boldsymbol{\Pi} \times \boldsymbol{\mathfrak{a}} - 2 m \boldsymbol{\mathfrak{t}}^{(1)} \,, \label{eq:dhw_v} \\
D_t \boldsymbol{\mathfrak{a}} &=& - \boldsymbol{D} \mathfrak{a}^0 - 2 \boldsymbol{\Pi} \times \boldsymbol{\mathfrak{v}} \,, \label{eq:dhw_a} \\
D_t \boldsymbol{\mathfrak{t}}^{(1)} &=& - \boldsymbol{D} \times \boldsymbol{\mathfrak{t}}^{(2)} - 2 \boldsymbol{\Pi} \mathfrak{s} + 2 m \boldsymbol{\mathfrak{v}} \,, \label{eq:dhw_t1} \\
D_t \boldsymbol{\mathfrak{t}}^{(2)} &=& \boldsymbol{D} \times \boldsymbol{\mathfrak{t}}^{(1)} + 2 \boldsymbol{\Pi} \mathfrak{p} \,,\label{eq:dhw_t2}
\end{eqnarray}
where the pseudodifferential operators read
\begin{eqnarray}
D_t &=& \partial_t + e \int_{-1/2}^{1/2} d\lambda \, \mathbf{E} (t, \mathbf{x}+\I \lambda \boldsymbol{\nabla}_{\mathbf{p}}) \boldsymbol{\nabla}_{\mathbf{p}} \,, \label{eq:Dt} \\
\boldsymbol{D}&=& \boldsymbol{\nabla} + e \int_{-1/2}^{1/2} d\lambda \, \mathbf{B} (t, \mathbf{x}+\I \lambda \boldsymbol{\nabla}_\mathbf{p}) \times \boldsymbol{\nabla}_\mathbf{p} \,,  \label{eq:Dx} \\
\boldsymbol{\Pi} &=& \mathbf{p} - \I e \int_{-1/2}^{1/2} d\lambda \, \lambda \mathbf{B} (t, \mathbf{x}+\I \lambda \boldsymbol{\nabla}_\mathbf{p}) \times \boldsymbol{\nabla}_\mathbf{p}.  \label{eq:Pi}
\end{eqnarray}    
By solving this system with the initial condition $W(\mathbf{x}, \mathbf{p}, t_\text{in}) = W_{\mathbf{A}=\mathbf{0}} (\mathbf{p})$, one can then extract the number density of the electrons via~\cite{blinne_gies_2014,blinne_strobel_2016,olugh_prd_2019,kohlfuerst_prd_2019,aleksandrov_kohlfuerst}
\begin{equation}
n^{(e^-)}_{\mathbf{p}} \equiv \sum_{s} n^{(e^-)}_{\mathbf{p},s} = \frac{1}{(2\pi)^3} \int d\mathbf{x} \, \mathfrak{e} (\mathbf{x}, \mathbf{p} - e\mathbf{A}_\text{out}, t_\text{out}) \,,
\label{eq:dhw_n_e}
\end{equation}
where
\begin{eqnarray}
\mathfrak{e} (\mathbf{x}, \mathbf{p}, t) &\equiv& \frac{m \big [ \mathfrak{s} (\mathbf{x}, \mathbf{p}, t) -\mathfrak{s}_{\mathbf{A}=\mathbf{0}} (\mathbf{p}) \big ]}{2p^0 (\mathbf{p})} \nonumber \\
{}&+& \frac{\mathbf{p} \big [
\boldsymbol{\mathfrak{v}} (\mathbf{x}, \mathbf{p}, t) - \boldsymbol{\mathfrak{v}}_{\mathbf{A}=\mathbf{0}} (\mathbf{p}) \big ]}{2p^0 (\mathbf{p})}\,.
\label{eq:expt}
\end{eqnarray}
Note that this distribution function involves summation over spin. The numerator in Eq.~\eqref{eq:expt} represents the phase-space energy density of the particles produced. The total number of pairs reads
\begin{eqnarray}
N &=& \int d\mathbf{p} \, n^{(e^-)}_{\mathbf{p}} =  \int d\mathbf{p} \, n^{(e^+)}_{\mathbf{p}} \nonumber \\
{}&=& \int \frac{d\mathbf{p}}{(2\pi)^3} \int d\mathbf{x} \, \mathfrak{e} (\mathbf{x}, \mathbf{p}, t_\text{out}) \,.
\end{eqnarray}
Next, we will consider the DHW formalism in the case of spatially uniform backgrounds.

%%%

\begin{widetext}
\subsection{Uniform external field. Wigner function via the {\it in} solutions}

Here we turn to the analysis of setups involving purely time-dependent external fields. Having described the Furry-picture formalism (Sec.~\ref{sec:furry}), we are now able to demonstrate how it is related to the DHW approach. To this end, we will calculate the Wigner function~\eqref{eq:wigner_function_def} by using the field-operator expansion~\eqref{eq:psi_x_in} in terms of the {\it in} solutions since the expectation value in Eq.~\eqref{eq:wigner_function_def} regards the {\it in} vacuum state. The Wigner function is obviously $\mathbf{x}$-independent, and taking into account the expressions~\eqref{eq:phi_plus_chi_1}--\eqref{eq:phi_minus_chi_2}, one derives the following representation:
\begin{equation}
W (\mathbf{p}, t) = -\frac{1}{2} \sum_{s} \Big [ {}_+\chi_{\mathbf{p} + e \mathbf{A}(t), s} (t) {}_+ \overline{\chi}_{\mathbf{p} + e \mathbf{A}(t), s} (t) - {}_-\chi_{-\mathbf{p} - e \mathbf{A}(t), s} (t) {}_- \overline{\chi}_{-\mathbf{p} - e \mathbf{A}(t), s} (t) \Big ] \,.
\label{eq:W_uniform}
\end{equation}
At $t = t_\text{in}$ we use Eqs.~\eqref{eq:chi_initial_cond_in_plus} and \eqref{eq:chi_initial_cond_in_minus} and obtain the zero-field expression~\eqref{eq:wigner_zero_field_W}.

The DHW system~\eqref{eq:dhw_s}--\eqref{eq:dhw_t2} involving 16 components is now significantly simplified. First, we notice that the functions $\mathfrak{p}$, $\mathfrak{v}^0$, $\mathfrak{a}^0$, and $\boldsymbol{\mathfrak{t}}^{(2)}$ vanish as they satisfy the homogeneous subsystem and equal zero at the initial time instant $t = t_\text{in}$. Then, the remaining ten DHW functions obey the system
\begin{eqnarray}
\big [ \partial_t + e \mathbf{E}(t) \boldsymbol{\nabla}_\mathbf{p} \big ] \mathfrak{s} &=& 2 \mathbf{p} \boldsymbol{\mathfrak{t}}^{(1)} \,, \label{eq:dhw_uni_s} \\
\big [ \partial_t + e \mathbf{E}(t) \boldsymbol{\nabla}_\mathbf{p} \big ] \boldsymbol{\mathfrak{v}} &=& -2 \mathbf{p} \times \boldsymbol{\mathfrak{a}} - 2 m \boldsymbol{\mathfrak{t}}^{(1)} \,, \label{eq:dhw_uni_v} \\
\big [ \partial_t + e \mathbf{E}(t) \boldsymbol{\nabla}_\mathbf{p} \big ] \boldsymbol{\mathfrak{a}} &=& -2 \mathbf{p} \times \boldsymbol{\mathfrak{v}} \,, \label{eq:dhw_uni_a} \\
\big [ \partial_t + e \mathbf{E}(t) \boldsymbol{\nabla}_\mathbf{p} \big ] \boldsymbol{\mathfrak{t}}^{(1)} &=& -2 \mathbf{p} \mathfrak{s} + 2 m \boldsymbol{\mathfrak{v}} \,. \label{eq:dhw_uni_t1}
\end{eqnarray}
The solution can be found in the form $\mathfrak{s} (\mathbf{p}, t) = \tilde{\mathfrak{s}} (\mathbf{p} + e \mathbf{A} (t), t)$. Then the system~\eqref{eq:dhw_uni_s}--\eqref{eq:dhw_uni_t1} reads
\begin{eqnarray}
\dot{\tilde{\mathfrak{s}}} &=& 2 \mathbf{q} \tilde{\boldsymbol{\mathfrak{t}}} \,, \label{eq:dhw_uni_s_tilde} \\
\dot{\tilde{\boldsymbol{\mathfrak{v}}}} &=& -2 \mathbf{q} \times \tilde{\boldsymbol{\mathfrak{a}}} - 2 m \tilde{\boldsymbol{\mathfrak{t}}} \,, \label{eq:dhw_uni_v_tilde} \\
\dot{\tilde{\boldsymbol{\mathfrak{a}}}} &=& -2 \mathbf{q} \times \tilde{\boldsymbol{\mathfrak{v}}} \,, \label{eq:dhw_uni_a_tilde} \\
\dot{\tilde{\boldsymbol{\mathfrak{t}}}} &=& -2 \mathbf{q} \tilde{\mathfrak{s}} + 2 m \tilde{\boldsymbol{\mathfrak{v}}} \,. \label{eq:dhw_uni_t1_tilde}
\end{eqnarray}
where the functions are evaluated at $\mathbf{p}$, $t$ and $\mathbf{q} = \mathbf{p} - e\mathbf{A}(t)$ in accordance with the definition of the kinetic momentum $\mathbf{q}$ introduced in Sec.~\ref{sec:qke}. We have also omitted the superscript of the vector function $\tilde{\boldsymbol{\mathfrak{t}}}^{(1)}$. The expression~\eqref{eq:W_uniform} in terms of the tilded components takes the form
\begin{equation}
\tilde{W} (\mathbf{p}, t) = W(\mathbf{p} - e\mathbf{A}(t),t) = -\frac{1}{2} \sum_{s} \Big [ {}_+\chi_{\mathbf{p},s} (t) {}_+ \overline{\chi}_{\mathbf{p}, s} (t) - {}_-\chi_{-\mathbf{p}, s} (t) {}_- \overline{\chi}_{-\mathbf{p}, s} (t) \Big ] \,.
\label{eq:W_uniform_tilde}
\end{equation}

Let us now establish the connection between the Wigner function and the $G$ matrices, i.e., the key object within the framework of Furry-picture quantization [see Eqs.~\eqref{eq:G_g} and \eqref{eq:g_chi}]. First, we express the {\it in} solutions in terms of the {\it out} ones according to
\begin{eqnarray}
{}_+\chi_{\mathbf{p},s} (t) &=& \sum_{s'} \Big [ g^* ({}_+| {}^+)_{\mathbf{p},s,s'} {}^+ \chi_{\mathbf{p}, s'} (t) + g^* ({}_+| {}^-)_{\mathbf{p},s,s'} {}^- \chi_{-\mathbf{p}, s'} (t) \Big ] \,, \label{eq:in_out_g_plus} \\
{}_-\chi_{\mathbf{p},s} (t) &=& \sum_{s'} \Big [ g^* ({}_-| {}^+)_{\mathbf{p},s,s'} {}^+ \chi_{-\mathbf{p}, s'} (t) + g^* ({}_-| {}^-)_{\mathbf{p},s,s'} {}^- \chi_{\mathbf{p}, s'} (t) \Big ] \label{eq:in_out_g_minus} \,.
\end{eqnarray}
Then, we substitute these expressions into Eq.~\eqref{eq:W_uniform} and utilize Eqs.~\eqref{eq:chi_initial_cond_out_plus} and \eqref{eq:chi_initial_cond_out_minus} at $t = t_\text{out}$. This allows one to extract the asymptotic values of the DHW functions in terms of the $G$ matrices. To simplify the expressions, we make use of the properties~\eqref{eq:G_matrix_orthonorm} and \eqref{eq:G_matrix_complete}. For instance,
\begin{eqnarray}
\tilde{\mathfrak{s}} (\mathbf{p}, t_\text{out}) &=& -\frac{2m}{q^0} + \frac{m}{q^0} \sum_{s, s'} \bigg [ \big | g({}_-| {}^+)_{-\mathbf{p},s,s'}\big |^2 + \big | g({}_+| {}^-)_{\mathbf{p},s,s'}\big |^2 \bigg ] \nonumber \\
{} &+& \sum_{s,s',s''} \frac{\mathbf{q} \boldsymbol{\sigma}_{s' s''}}{q^0} \Big [ g({}_+| {}^-)_{\mathbf{p},s,s'} g^*({}_+| {}^+)_{\mathbf{p},s,s''} + g({}_+| {}^+)_{\mathbf{p},s,s'} g^*({}_+| {}^-)_{\mathbf{p},s,s''} \Big ] \,, \label{eq:dhw_s_tilde}\\
\tilde{\boldsymbol{\mathfrak{v}}} (\mathbf{p}, t_\text{out}) &=& -\frac{2\mathbf{q}}{q^0} + \frac{\mathbf{q}}{q^0} \sum_{s, s'} \bigg [ \big | g({}_-| {}^+)_{-\mathbf{p},s,s'}\big |^2 + \big | g({}_+| {}^-)_{\mathbf{p},s,s'}\big |^2 \bigg ] \nonumber \\
{} &-& \sum_{s,s',s''} \bigg [ \boldsymbol{\sigma}_{s's''} - \frac{\mathbf{q} (\mathbf{q} \boldsymbol{\sigma}_{s' s''})}{q^0 (q^0 + m)} \bigg ] \Big [ g({}_+| {}^-)_{\mathbf{p},s,s'} g^*({}_+| {}^+)_{\mathbf{p},s,s''} + g({}_+| {}^+)_{\mathbf{p},s,s'} g^*({}_+| {}^-)_{\mathbf{p},s,s''} \Big ] \,, \label{eq:dhw_v_tilde}
\end{eqnarray}
where $\mathbf{q} = \mathbf{p} - e\mathbf{A} (t_\text{out})$. By using the definition~\eqref{eq:expt} and Eqs.~\eqref{eq:dhw_s_tilde} and \eqref{eq:dhw_v_tilde}, we now easily obtain
\begin{eqnarray}
\tilde{\mathfrak{e}} (\mathbf{p}, t_\text{out}) &=& \frac{1}{2} \sum_{s, s'} \bigg [ \big | g({}_-| {}^+)_{-\mathbf{p},s,s'}\big |^2 + \big | g({}_+| {}^-)_{\mathbf{p},s,s'}\big |^2 \bigg ] \nonumber \\
{} &=& \frac{1}{2} \frac{(2\pi)^3}{V} \, \big [ n^{(e^-)}_{\mathbf{p}} + n^{(e^+)}_{-\mathbf{p}} \big ] = \frac{(2\pi)^3}{V} \, n^{(e^-)}_{\mathbf{p}} \,,
\label{eq:dhw_e_tilde}
\end{eqnarray}
so we recover Eq.~\eqref{eq:dhw_n_e}. One can also explicitly confirm that the DHW functions $\mathfrak{p}$, $\mathfrak{v}^0$, $\mathfrak{a}^0$, and $\boldsymbol{\mathfrak{t}}^{(2)}$ vanish. Finally, we note that while the DHW components are introduced directly via the Wigner function~\eqref{eq:wigner_function_def} and do not imply any specific choice of the basis, the complex phases of the $G$ matrices depend on the choice of the asymptotic states~\eqref{eq:chi_initial_cond_in_plus}--\eqref{eq:chi_initial_cond_out_minus}. The coefficients in the relations~\eqref{eq:dhw_s_tilde} and \eqref{eq:dhw_v_tilde} are particularly simple if the {\it in} and {\it out} states have vanishing phases at $t=t_\text{in}$ and $t=t_\text{out}$ as we defined in Eqs.~\eqref{eq:chi_initial_cond_in_plus}--\eqref{eq:chi_initial_cond_out_minus}. The physical number of particles~\eqref{eq:dhw_e_tilde} does not depend on these phases.

Next we will connect the DHW formalism with the QKE approach and show the equivalence of the systems~\eqref{eq:dhw_uni_s_tilde}--\eqref{eq:dhw_uni_t1_tilde} and \eqref{eq:system_simple_f_s}--\eqref{eq:system_simple_v_v}.

%%%

\subsection{Wigner function via the adiabatic basis. Relation to the QKEs}

To establish a direct connection between the DHW formalism and the QKEs, we will now decompose the Heisenberg field operator in terms of the adiabatic basis [see Eq.~\eqref{eq:psi_adiabatic}] and compute the Wigner function. Since the explicit form of the adiabatic functions is known, the only difficulty is to calculate the {\it in}-vacuum expectation values of the adiabatic creation and annihilation operators. The latter were already found in Eqs.~\eqref{eq:vac_mean_2}, so the remaining part of the computations is straightforward. Namely, by substituting Eq.~\eqref{eq:psi_adiabatic} into Eqs.~\eqref{eq:C_def} and \eqref{eq:W_op_s}, we find
\begin{eqnarray}
\hat{W} (\mathbf{x}, \mathbf{p}, t) &=& W_{\mathbf{A}=\mathbf{0}} (\mathbf{p}) + 2 \sum_{s', s''} \int \! d\mathbf{p}' \, \Big \{ a_{-\mathbf{p}' + \mathbf{p} + e\mathbf{A} (t), s''}^\dagger (t) \, a_{\mathbf{p}' + \mathbf{p} + e\mathbf{A} (t), s'} (t) u_{\mathbf{p}' + \mathbf{p},s'} \overline{u}_{-\mathbf{p}'+\mathbf{p},s''} \, \mathrm{e}^{2\I \mathbf{p}' \mathbf{x}} \nonumber \\
{}&+& b_{\mathbf{p}' - \mathbf{p} - e\mathbf{A} (t), s''} (t) \, a_{\mathbf{p}' + \mathbf{p} + e\mathbf{A} (t), s'} (t) u_{\mathbf{p}' + \mathbf{p},s'} \overline{v}_{-\mathbf{p}'+\mathbf{p},s''} \, \mathrm{e}^{2\I \mathbf{p}' \mathbf{x}} \nonumber \\
{}&+& a_{\mathbf{p}' + \mathbf{p} + e\mathbf{A} (t), s''}^\dagger (t) \, b_{\mathbf{p}' - \mathbf{p} - e\mathbf{A} (t), s'}^\dagger (t) v_{-\mathbf{p}' + \mathbf{p},s'} \overline{u}_{\mathbf{p}'+\mathbf{p},s''} \, \mathrm{e}^{-2\I \mathbf{p}' \mathbf{x}} \nonumber \\
{}&-& b_{\mathbf{p}' - \mathbf{p} - e\mathbf{A} (t), s'}^\dagger (t) \, b_{-\mathbf{p}' - \mathbf{p} - e\mathbf{A} (t), s''} (t) v_{-\mathbf{p}' + \mathbf{p},s'} \overline{v}_{\mathbf{p}'+\mathbf{p},s''} \, \mathrm{e}^{-2\I \mathbf{p}' \mathbf{x}} \Big \} .
\end{eqnarray}
By evaluating then the {\it in}-vacuum expectation value using Eqs.~\eqref{eq:vac_mean_2}, we obtain the following expression for the $\mathbf{x}$-independent Wigner function in terms of the QKE components:
\begin{eqnarray}
W (\mathbf{p}, t) &=& \big [ 1 - 2f(\mathbf{p} + e \mathbf{A}(t), t) \big ] W_{\mathbf{A}=\mathbf{0}} (\mathbf{p}) \nonumber \\
{}&+& \sum_{s',s''} \big [ \mathbf{f} (\mathbf{p} + e \mathbf{A}(t), t) \boldsymbol{\sigma}_{s's''} \big ] \big [ u_{\mathbf{p},s'} \overline{u}_{\mathbf{p},s''} + v_{\mathbf{p},s'} \overline{v}_{\mathbf{p},s''} \big ] \nonumber \\
{}&+& \sum_{s',s''} \big [ \mathbf{u} (\mathbf{p} + e \mathbf{A}(t), t) \boldsymbol{\sigma}_{s's''} \big ] \big [ u_{\mathbf{p},s'} \overline{v}_{\mathbf{p},s''} + v_{\mathbf{p},s'} \overline{u}_{\mathbf{p},s''} \big ] \nonumber \\
{}&+& \I \sum_{s',s''} \big [ \mathbf{v} (\mathbf{p} + e \mathbf{A}(t), t) \boldsymbol{\sigma}_{s's''} \big ] \big [ u_{\mathbf{p},s'} \overline{v}_{\mathbf{p},s''} - v_{\mathbf{p},s'} \overline{u}_{\mathbf{p},s''} \big ] \,.
\label{eq:W_qke_general}
\end{eqnarray}
By calculating the traces according to Eqs.~\eqref{eq:wigner_coeff} and taking into account the explicit form of the bispinors~\eqref{eq:u_explicit}--\eqref{eq:v_explicit}, we find
\begin{eqnarray}
\tilde{\mathfrak{s}} - \tilde{\mathfrak{s}}_{\mathbf{A}=\mathbf{0}} &=& \frac{4}{q^0} \big [ m f - (\mathbf{q} \mathbf{u}) \big ] \,, \label{eq:rel_s}\\
\tilde{\boldsymbol{\mathfrak{v}}} - \tilde{\boldsymbol{\mathfrak{v}}}_{\mathbf{A}=\mathbf{0}} &=& \frac{4}{q^0} \bigg [ q^0 \mathbf{u} + \mathbf{q} f - \frac{\mathbf{q} (\mathbf{q} \mathbf{u})}{q^0+m} \bigg ] \,, \label{eq:rel_v}\\
\tilde{\boldsymbol{\mathfrak{a}}} &=& -\frac{4}{q^0} \bigg [ m \mathbf{f} - (\mathbf{q} \times \mathbf{v}) + \frac{\mathbf{q} (\mathbf{q} \mathbf{f})}{q^0+m} \bigg ] \,, \label{eq:rel_a} \\
\tilde{\boldsymbol{\mathfrak{t}}} &=& -\frac{4}{q^0} \bigg [ m \mathbf{v} - (\mathbf{q} \times \mathbf{f}) + \frac{\mathbf{q} (\mathbf{q} \mathbf{v})}{q^0+m} \bigg ] \,. \label{eq:rel_t1} 
\end{eqnarray}
Here all of the functions are evaluated at $\mathbf{p}$, $t$ and $\mathbf{q} = \mathbf{p} - e\mathbf{A}(t)$. We remind the reader that $q^0 = \omega (\mathbf{p}, t)$. Note that $\tilde{\mathfrak{s}}$ and $\tilde{\boldsymbol{\mathfrak{v}}}$ are expressed only in terms of $f$ and $\mathbf{u}$, while $\tilde{\boldsymbol{\mathfrak{a}}}$ and $\tilde{\boldsymbol{\mathfrak{t}}}$ are determined only by $\mathbf{f}$ and $\mathbf{v}$. The inverse transformation is given by
\begin{eqnarray}
f &=& \frac{1}{4q^0} \big [ m(\tilde{\mathfrak{s}} - \tilde{\mathfrak{s}}_{\mathbf{A}=\mathbf{0}} ) + \mathbf{q} (\tilde{\boldsymbol{\mathfrak{v}}} - \tilde{\boldsymbol{\mathfrak{v}}}_{\mathbf{A}=\mathbf{0}}) \big ] \,, \label{eq:rel_qke_f} \\
\mathbf{u} &=& -\frac{1}{4q^0} \bigg [ \mathbf{q} \tilde{\mathfrak{s}} - q^0 \tilde{\boldsymbol{\mathfrak{v}}} + \frac{\mathbf{q} (\mathbf{q} \tilde{\boldsymbol{\mathfrak{v}}})}{q^0+m} \bigg ] \,, \label{eq:rel_qke_u} \\
\mathbf{f} &=& - \frac{1}{4q^0} \bigg [ m\tilde{\boldsymbol{\mathfrak{a}}} + (\mathbf{q} \times \tilde{\boldsymbol{\mathfrak{t}}}) + \frac{\mathbf{q} (\mathbf{q} \tilde{\boldsymbol{\mathfrak{a}}}  )}{q^0+m} \bigg ] \,, \label{eq:rel_qke_f_v} \\
\mathbf{v} &=& - \frac{1}{4q^0} \bigg [ m\tilde{\boldsymbol{\mathfrak{t}}} + (\mathbf{q} \times \tilde{\boldsymbol{\mathfrak{a}}}) + \frac{\mathbf{q} (\mathbf{q} \tilde{\boldsymbol{\mathfrak{t}}})}{q^0+m} \bigg ] \,. \label{eq:rel_qke_v} 
\end{eqnarray}
In Eq.~\eqref{eq:rel_qke_u} the zero-field contributions exactly canceled each other out. The initial values of the DHW and QKE functions are in agreement. It is now also obvious that $\tilde{\mathfrak{e}} (\mathbf{p}, t) = 2f (\mathbf{p}, t)$, which is in accordance with Eqs.~\eqref{eq:number_density_qke_el} and \eqref{eq:dhw_e_tilde}.

We also point out that the DHW system~\eqref{eq:dhw_uni_s_tilde}--\eqref{eq:dhw_uni_t1_tilde} is often employed in a different form as discussed in Appendix~\ref{app_dhw_alt}. This alternative system is utilized, e.g., in Refs.~\cite{blinne_strobel_2016,olugh_prd_2019,hu_prd_2023,hu_arxiv_2024}.

We have explicitly discovered the connection between the ten DHW components and ten QKE functions, which is one of the main findings of the present paper. Next, let us discuss the physical interpretation of the spin-resolved particle densities, which can be extracted from the two systems.

%%%

\subsection{Physical interpretation of the spin effects. Helicity states} \label{sec:helicity}

While we already showed how one should combine the DHW functions to obtain the momentum spectra of the particles summed over spin [see Eq.~\eqref{eq:dhw_e_tilde}], it may be not yet clear how to obtain the spin-resolved distributions, i.e. the DHW analog of Eq.~\eqref{eq:number_density_qke_el} should be derived. Moreover, the physical interpretation of the anisotropy in Eq.~\eqref{eq:number_density_qke_el} should also be clarified.

Although one can, in principle, experimentally measure the densities~\eqref{eq:number_density_qke_el} if the electrons are projected onto the corresponding states~\eqref{eq:u_explicit}, from the physical viewpoint, these states may not be relevant ones. While it is shown in Sec.~\ref{sec:qke_basis} that a {\it constant} rotation of the bispinor basis can hardly simplify the computations, one can easily perform a necessary transformation using the {\it final} values of the QKE functions. Since we always fix the momentum of the particles produced, it is the particle's helicity that represents a relevant additional observable as the helicity operator commutes with the Hamiltonian. To construct the corresponding eigenfunctions, one has to perform the following unitary transformation:
\begin{equation}
u_{\mathbf{p},s} = \alpha^{(\text{L})}_{\mathbf{p},s} \, u_\mathbf{p}^{(\text{L})} + \alpha^{(\text{R})}_{\mathbf{p},s} \, u_\mathbf{p}^{(\text{R})} \,, \label{eq:hel_u}
\end{equation}
where
\begin{equation}
\alpha^{(\text{L})}_{\mathbf{p},-} = \alpha^{(\text{R})}_{\mathbf{p},+} = \frac{1}{\sqrt{2}} \sqrt{\frac{|\mathbf{p}|-p_z}{|\mathbf{p}|}} \,, \qquad \alpha^{(\text{L})}_{\mathbf{p},+} = - \big [ \alpha^{(\text{R})}_{\mathbf{p},-} \big ]^* = - \frac{1}{\sqrt{2}} \frac{p_x-\I p_y}{\sqrt{|\mathbf{p}| (|\mathbf{p}| - p_z)}}\,.
\end{equation}
One can directly verify that $u_\mathbf{p}^{(\text{L})}$ and $u_\mathbf{p}^{(\text{R)}}$ are the eigenvectors of the helicity operator $(\boldsymbol{\Sigma} \mathbf{p})/|\mathbf{p}|$. The transformation of the bispinors $v_{\mathbf{p},s}$ is exactly the same. To obtain the helicity-resolved densities, one has to perform the corresponding Bogoliubov transformation of the creation/annihilation operators:
\begin{equation}
a_\mathbf{p}^{(\text{L})} (t) = \sum_{s=\pm} \alpha^{(\text{L})}_{\mathbf{q},s} \, a_{\mathbf{p},s} (t) \,, \qquad a_\mathbf{p}^{(\text{R})} (t) = \sum_{s=\pm} \alpha^{(\text{R})}_{\mathbf{q},s} \, a_{\mathbf{p},s} (t) \,, \label{eq:bogoliubov}
\end{equation}
where $\mathbf{q} = \mathbf{p} - e \mathbf{A} (t)$, so the coefficients depend on time. The electron density functions corresponding to a negative helicity (L) and to a positive helicity (R) are then given by
\begin{eqnarray}
f^{(e^-\text{L})}(\mathbf{p}, t) &\equiv& \frac{(2\pi)^3}{V} \big \langle 0, \text{in} \big | \big [ a_\mathbf{p}^{(\text{L})} (t)  \big ]^\dagger a_\mathbf{p}^{(\text{L})} (t) \big | 0,\text{in} \big \rangle = f(\mathbf{p}, t) - \frac{\mathbf{q} \mathbf{f} (\mathbf{p}, t)}{|\mathbf{q}|} \,, \label{eq:fL}\\
f^{(e^-\text{R})}(\mathbf{p}, t) &\equiv& \frac{(2\pi)^3}{V} \big \langle 0, \text{in} \big | \big [ a_\mathbf{p}^{(\text{R})} (t)  \big ]^\dagger a_\mathbf{p}^{(\text{R})} (t) \big | 0,\text{in} \big \rangle = f(\mathbf{p}, t) + \frac{\mathbf{q} \mathbf{f} (\mathbf{p}, t)}{|\mathbf{q}|} \,. \label{eq:fR}
\end{eqnarray}
The final densities of the electrons produced read
\begin{eqnarray}
\frac{(2\pi)^3}{V} n^{(e^-\text{L})}_\mathbf{p} &=& f^{(e^-\text{L})}(\mathbf{p}, t_\text{out}) \,, \label{eq:n_eL}\\
\frac{(2\pi)^3}{V} n^{(e^-\text{R})}_\mathbf{p} &=& f^{(e^-\text{R})}(\mathbf{p}, t_\text{out}) \,. \label{eq:n_eR}
\end{eqnarray}
The sum over $s$ in Eq.~\eqref{eq:number_density_qke_el} obviously coincides with a sum of these two quantities. For the positrons produced, the analogous expressions have the form
\begin{eqnarray}
\frac{(2\pi)^3}{V} n^{(e^+\text{L})}_{\mathbf{p}} &=& f^{(e^-\text{R})}(-\mathbf{p}, t_\text{out}) \,, \label{eq:n_pL}\\
\frac{(2\pi)^3}{V} n^{(e^+\text{R})}_{\mathbf{p}} &=& f^{(e^-\text{L})}(-\mathbf{p}, t_\text{out}) \,. \label{eq:n_pR}
\end{eqnarray}

By using the relation~\eqref{eq:rel_qke_f_v}, one can immediately refer the helicity-resolved distributions to the DHW functions, so the Bogoliubov transformation within the QKE approach allows one to explicitly construct the necessary observables also in the DHW framework. One straightforwardly obtains
\begin{equation}
f^{(e^-\text{L/R})}(\mathbf{p}, t) = f(\mathbf{p}, t) \pm \frac{\mathbf{q} \tilde{\boldsymbol{\mathfrak{a}}} (\mathbf{p}, t)}{4|\mathbf{q}|} \,.
\end{equation}
We also note that the $G$ matrices within the Furry-picture formalism can also be easily transformed in order to describe the helicity states.

Although the helicity-resolved observables can be acquired by means of the Bogoliubov transformation as was done above, it is also possible to obtain them directly within the Wigner-function approach. To this end, we will first introduce the adiabatic densities corresponding to given values of $s$ analogous to those defined in Eqs.~\eqref{eq:fL}--\eqref{eq:fR}:
\begin{equation}
f^{(e^-)}_s(\mathbf{p}, t) \equiv \frac{(2\pi)^3}{V} \big \langle 0, \text{in} \big | a^\dagger_{\mathbf{p},s} (t) a_{\mathbf{p},s} (t) \big | 0,\text{in} \big \rangle \,. \label{eq:fs}
\end{equation}
Let us note then that one can straightforwardly prove the following identity:
\begin{equation}
\tilde{\mathfrak{e}} (\mathbf{p}, t) = \mathrm{Tr} \ \bigg \{ \frac{m \mathrm{I} + \boldsymbol{\gamma} \mathbf{q}}{2q^0} \big [ \tilde{W} (\mathbf{p}, t) - \tilde{W}_{\mathbf{A}=\mathbf{0}} (\mathbf{p}, t) \big ]  \bigg \} \,,
\end{equation}
where $\tilde{W}_{\mathbf{A}=\mathbf{0}} (\mathbf{p}, t) = W_{\mathbf{A}=\mathbf{0}}(\mathbf{p} - e\mathbf{A}(t))$ and $\mathbf{q} = \mathbf{p} - e \mathbf{A}(t)$. According to Eq.~\eqref{eq:dhw_e_tilde}, the function $\tilde{\mathfrak{e}} (\mathbf{p}, t)$ is a sum of the electron and positron contributions with both values of $s$ (multiplied by a factor of $1/2$). In order to obtain the partial term corresponding to the density~\eqref{eq:fs} for a given $s$, one should apply the corresponding projection operator:
\begin{equation}
f^{(e^-)}_s(\mathbf{p}, t) = \mathrm{Tr} \ \bigg \{ \frac{m \mathrm{I} + \boldsymbol{\gamma} \mathbf{q}}{q^0} \, \hat{\mathcal{P}}^{(e^-)}_s \big [ \tilde{W} (\mathbf{p}, t) - \tilde{W}_{\mathbf{A}=\mathbf{0}} (\mathbf{p}, t) \big ]  \bigg \} \,.
\label{eq:fs_proj}
\end{equation}
The explicit form of $\hat{\mathcal{P}}^{(e^-)}_s$ is derived in Appendix~\ref{app_dhw_proj}, and the final prescription reads
\begin{equation}
f^{(e^-)}_s(\mathbf{p}, t) = \mathrm{Tr} \ \Big \{ \gamma^0 u_{\mathbf{q},s} u^\dagger_{\mathbf{q},s} \big [ \tilde{W} (\mathbf{p}, t) - \tilde{W}_{\mathbf{A}=\mathbf{0}} (\mathbf{p}, t) \big ]  \Big \} \,.
\label{eq:fs_uud}
\end{equation}
Note that this quantity is real since $W^\dagger = \gamma^0 W \gamma^0$ as was pointed out in Sec.~\ref{sec:dhw_gen} (cf.~Ref.~\cite{blinne_strobel_2016}, where similar expressions were employed without $\gamma^0$). We underline here that the definition of the Wigner function~\eqref{eq:wigner_function_def} involves the field operators themselves, i.e. $W$ and $\tilde{W}$ are independent of whether one uses any decomposition of $\psi$. In other words, the Wigner-function formalism is inherently basis-independent, while it is the projection matrix in Eq.~\eqref{eq:fs_uud} that yields the particle density of interest. By evaluating the trace and using Eq.~\eqref{eq:W_qke_general}, we find
\begin{equation}
f^{(e^-)}_s(\mathbf{p}, t) =  f (\mathbf{p}, t) - (\mathrm{sign} \, s) \, f_z (\mathbf{p}, t) \,.
\end{equation}
This result coincides with Eq.~\eqref{eq:number_density_qke_el}.

To extract other particle densities, one should simply replace $u_{\mathbf{q},s} u^\dagger_{\mathbf{q},s}$ in Eq.~\eqref{eq:fs_uud} with the relevant projection matrix. For instance, to obtain a helicity-resolved density, one evaluates
\begin{eqnarray}
f^{(e^-\text{L/R})}(\mathbf{p}, t) &=& \mathrm{Tr} \ \Big \{ \gamma^0 u^{(\text{L/R})}_{\mathbf{q}} \big [ u^{(\text{L/R})}_{\mathbf{q}} \big ]^\dagger \big [ \tilde{W} (\mathbf{p}, t) - \tilde{W}_{\mathbf{A}=\mathbf{0}} (\mathbf{p}, t) \big ]  \Big \} \nonumber \\
{}&=& f(\mathbf{p}, t) \mp \frac{\mathbf{q} \mathbf{f} (\mathbf{p}, t)}{|\mathbf{q}|}  \label{eq:fL_uud} \,,
\end{eqnarray}
which confirms Eqs.~\eqref{eq:fL} and \eqref{eq:fR}. The helicity-resolved distributions can be then computed numerically~\cite{aleksandrov_kudlis_prd}.

\end{widetext}

%%%%%%%%%%%%%%%%%%%%%%%%%%%%%%%%%%%%%%%%%%%%%%%%%%%%%%%%%

\section{Conclusion} \label{sec:conclusion}

In this study, we revisited the kinetic approach to describing vacuum pair production in strong electric backgrounds. We examined the techniques based on the QKEs and the DHW formalism and demonstrated their equivalence. To this end, we first considered the framework of Furry-picture quantization, where the observable number densities of the particles produced are expressed in terms of the one-particle transition amplitudes. Along the same lines, one can quantize the electron-positron field using the basis of the adiabatic Hamiltonian eigenfunctions. Utilizing this approach, we derived first the QKEs and showed that the correct form of the equations is different to what was known in the literature. Second, we employed the adiabatic basis to directly compute the Wigner function, which allowed us to explicitly connect the kinetic functions involved in the DHW and QKE techniques.

A practical significance of the one-to-one correspondences derived in this study was already demonstrated: it is very convenient to switch from one formalism to another on the level of the final expressions without performing all of the intermediate steps. Furthermore, a unified picture involving the available theoretical frameworks also provides a coherent set of tools from the viewpoint of numerical computations.

Finally, we addressed the effect of spin anisotropy, which appears in the particle distributions in the case of rotating external fields. We constructed the number densities corresponding to the states with well-defined helicity by means of both direct Bogoliubov transformation and projection method within the DHW formalism. A numerical implementation of the QKE system for the analysis of the helicity effects was recently performed in Ref.~\cite{aleksandrov_kudlis_prd}.

Our study is expected to illuminate the fundamental aspects of the kinetic description of vacuum pair creation and to provide a solid theoretical basis for computing the observable spectra of the particles produced.

%%%%%%%%%%%%%%%%%%%%%%%%%%%%%%%%%%%%%%%%%%%%%%%%%%%%%%%%%

\acknowledgments

The study was funded by the Russian Science Foundation, Project No.~23-72-01068. The analysis of the Dirac-Heisenberg-Wigner formalism in Sec.~III was supported by the Icelandic Research Fund (Ranns\'oknasj\'o{\dh}ur, Grant No.~2410550).

%%%%%%%%%%%%%%%%%%%%%%%%%%%%%%%%%%%%%%%%%%%%%%%%%%%%%%%%%

\appendix
\section{Reduction of the QKEs in the case of linear polarization} \label{app:LP}

Here we present a proof of the statement~\eqref{eq:LP_statement}. First, let us define
\begin{equation}
\boldsymbol{\mathcal{E}} = 2(\boldsymbol{\mu}_1 \times \mathbf{e}) - \dot{\mathbf{e}} \,.
\label{eq:E_def}
\end{equation}
The explicit form of this vector reads
\begin{eqnarray}
\boldsymbol{\mathcal{E}} &=& \frac{q^0}{\sqrt{q_0^2 - (\mathbf{q} \mathbf{n})^2}} \Bigg \{ \dot{\mathbf{n}} + \frac{(\mathbf{q}\dot{\mathbf{n}})}{q_0^2 - (\mathbf{q} \mathbf{n})^2} \nonumber \\
{}&\times& \bigg [ (\mathbf{q} \mathbf{n}) \mathbf{n} 
 - \frac{q^0}{q^0 + m} \, \mathbf{q} \bigg ] \Bigg \} \,.
\label{eq:E_explicit}
\end{eqnarray}
Since the particle momentum $\mathbf{p}$ governing $\mathbf{q} = \mathbf{p} - e\mathbf{A}$ is independent of the external field, from Eq.~\eqref{eq:E_explicit} it follows that
\begin{eqnarray}
\dot{\mathbf{n}} = \mathbf{0} \quad \Longleftrightarrow \quad \boldsymbol{\mathcal{E}} = \mathbf{0} \,.
\end{eqnarray}

Let us first assume that $\mathbf{f}$, $\mathbf{u}_\perp$, and $\mathbf{v}_\perp$ vanish and show that $\boldsymbol{\mathcal{E}} = \mathbf{0}$. The nonzero longitudinal projections of $\mathbf{u}$ and $\mathbf{v}$ together with $f$ obey Eqs.~\eqref{eq:qke_LP}. Since $\mathbf{u}_\perp = 0$, we have $\dot{\mathbf{u}} = \dot{u}_\parallel \mathbf{e}+u_\parallel \dot{\mathbf{e}}$, and the similar relation holds also for $\dot{\mathbf{v}}$. From Eqs.~\eqref{eq:system_simple_u_v} and \eqref{eq:system_simple_v_v} we obtain
\begin{equation}
\boldsymbol{\mathcal{E}} u_\parallel = \boldsymbol{\mathcal{E}} v_\parallel = \mathbf{0} \,.
\end{equation}
Thus, $\boldsymbol{\mathcal{E}} = \mathbf{0}$, which was to be demonstrated.

Let us now assume $\boldsymbol{\mathcal{E}} = \mathbf{0}$ and prove that $\mathbf{f}$, $\mathbf{u}_\perp$, and $\mathbf{v}_\perp$ vanish. For $\boldsymbol{\mathcal{E}} = \mathbf{0}$, the system~\eqref{eq:system_simple_f_s}--\eqref{eq:system_simple_v_v} can be rewritten in the form
\begin{eqnarray}
\begin{aligned}
\dot{f} &= - 2 \mu_2 u_\parallel \,, \\
\dot{f}_\parallel &= 0 \,, \\
\dot{u}_\parallel &= \mu_2 (2f-1) + 2\omega v_\parallel \,, \\
\dot{v}_\parallel &= -2\omega u_\parallel \,, \\
\dot{\mathbf{f}}_\perp &= 2 (\boldsymbol{\mu}_1 \times \mathbf{f}_\perp ) - 2 (\boldsymbol{\mu}_2 \times \mathbf{v}_\perp ) \,, \\
\dot{\mathbf{u}}_\perp &= 2 (\boldsymbol{\mu}_1 \times \mathbf{u}_\perp ) + 2 \omega \mathbf{v}_\perp \,, \\
\dot{\mathbf{v}}_\perp &= 2 (\boldsymbol{\mu}_1 \times \mathbf{v}_\perp ) - 2 (\boldsymbol{\mu}_2 \times \mathbf{f}_\perp ) - 2\omega \mathbf{u}_\perp \,.
\end{aligned}
\end{eqnarray}
Given the initial conditions, the last three equations yield $\mathbf{f}_\perp = \mathbf{u}_\perp = \mathbf{v}_\perp = \mathbf{0}$ and the second equation leads to $f_\parallel = 0$, which completes the proof. The resulting system is exactly that displayed in Eqs.~\eqref{eq:qke_LP}.

%%%%%%%%%%%%%%%%%%%%%%%%%%%%%%%%%%%%%%%%%%%%%%%%%%%%%%%%%

\section{Alternative form of the DHW system} \label{app_dhw_alt}

Let us now derive an alternative form of the DHW system~\eqref{eq:dhw_uni_s_tilde}--\eqref{eq:dhw_uni_t1_tilde}. Instead of the functions $\tilde{\mathfrak{s}}$ and $\tilde{\boldsymbol{\mathfrak{v}}}$ we will employ $f$ defined above and a new vector function $\boldsymbol{\mathfrak{w}}$,
\begin{eqnarray}
f &=& \frac{1}{2}+\frac{m \tilde{\mathfrak{s}} + \mathbf{q} \tilde{\boldsymbol{\mathfrak{v}}}}{4q^0} \,, \label{eq:new_f} \\
\boldsymbol{\mathfrak{w}} &=& \tilde{\boldsymbol{\mathfrak{v}}} - \frac{\mathbf{q}}{q_0^2} \, (\mathbf{q} \tilde{\boldsymbol{\mathfrak{v}}}) - \frac{m\mathbf{q}}{q_0^2} \, \tilde{\mathfrak{s}} \,. \label{eq:new_w}
\end{eqnarray}
Both of these functions vanish at $t = t_\text{in}$. The inverse transformation has the form
\begin{eqnarray}
\tilde{\mathfrak{s}} &=& -\frac{2m}{q^0} \, (1-2f) - \frac{\mathbf{q} \boldsymbol{\mathfrak{w}}}{m} \,, \label{eq:new_s} \\
\tilde{\boldsymbol{\mathfrak{v}}} &=& \boldsymbol{\mathfrak{w}} - \frac{2\mathbf{q}}{q^0} \, (1-2f) \,. \label{eq:new_v}
\end{eqnarray}
One can straightforwardly demonstrate that the DHW system takes the following form:
\begin{eqnarray}
\dot{f} &=& \frac{e}{4q^0} \, (\mathbf{E} \boldsymbol{\mathfrak{w}}) \,, \label{eq:dhw_new_f} \\
\dot{\boldsymbol{\mathfrak{w}}} &=& -2 \mathbf{q} \times \tilde{\boldsymbol{\mathfrak{a}}} - 2 m \tilde{\boldsymbol{\mathfrak{t}}} \nonumber \\
{}&+& \frac{2e}{q^0} \bigg [ \mathbf{E} - \frac{(\mathbf{q} \mathbf{E}) \mathbf{q}}{q_0^2}\bigg ] \big ( 1 - 2f \big) - \frac{e(\mathbf{E} \boldsymbol{\mathfrak{w}})\mathbf{q}}{q_0^2}\,, \label{eq:dhw_new_w} \\
\dot{\tilde{\boldsymbol{\mathfrak{a}}}} &=& -2 \mathbf{q} \times \boldsymbol{\mathfrak{w}} \,, \label{eq:dhw_new_a_tilde} \\
\dot{\tilde{\boldsymbol{\mathfrak{t}}}} &=& \frac{2}{m} \big [ m^2 \boldsymbol{\mathfrak{w}} + \mathbf{q} (\mathbf{q} \boldsymbol{\mathfrak{w}}) \big ] \,. \label{eq:dhw_new_t1_tilde}
\end{eqnarray}
For completeness, we also present the function $\boldsymbol{\mathfrak{w}}$ in terms of the QKE components (in fact, it is expressed only via $\mathbf{u}$) and the inverse relation:
\begin{eqnarray}
\boldsymbol{\mathfrak{w}} &=& \frac{4}{q^0} \bigg [ q^0 \mathbf{u} - \frac{\mathbf{q} (\mathbf{q} \mathbf{u})}{q^0+m} \bigg ] \,, \label{eq:rel_w} \\
\mathbf{u} &=& \frac{1}{4m} \bigg [ m \boldsymbol{\mathfrak{w}} + \frac{\mathbf{q} (\mathbf{q} \mathbf{\boldsymbol{\mathfrak{w}}})}{q^0+m} \bigg ] \,. \label{eq:rel_u}
\end{eqnarray}

%%%%%%%%%%%%%%%%%%%%%%%%%%%%%%%%%%%%%%%%%%%%%%%%%%%%%%%%%

\begin{widetext}
\section{Spin projections of the Wigner function} \label{app_dhw_proj}

Here we will derive the operator $\hat{\mathcal{P}}^{(e^-)}_s$, which allows one to extract the number density of the electrons produced with a given value of $s$ according to Eq.~\eqref{eq:fs_proj}. The main idea is to replace $\psi$ and $\overline{\psi}$ in Eq.~\eqref{eq:C_def} with their projections onto the corresponding subspace of $u$-bispinors with index $s$. The projection operator has the form
\begin{equation}
\hat{P}^{(e^-)}_s \psi (t, \mathbf{x}) = \int \! d\mathbf{x}' \int \! \frac{d\mathbf{p}'}{(2\pi)^3} \, \mathrm{e}^{\I \mathbf{p}' (\mathbf{x} - \mathbf{x}')} u_{\mathbf{p}' - e \mathbf{A}(t),s} u^\dagger_{\mathbf{p}' - e \mathbf{A}(t),s} \psi (t, \mathbf{x}') \,.
\label{eq:Ppsi}
\end{equation}
One can easily demonstrate that $\hat{P}^{(e^-)}_s$ is Hermitian and idempotent. In Eq.~\eqref{eq:C_def} we now substitute $\psi$ for~\eqref{eq:Ppsi}. Accordingly, the necessary projection of the Wigner function is given by
\begin{equation*}
\hat{\mathcal{P}}^{(e^-)}_s W (\mathbf{p}, t) = - \frac{1}{2} \int \! d\mathbf{s} \, \mathrm{e}^{-\I [\mathbf{p} + e \mathbf{A} (t)] \mathbf{s}} \big \langle 0, \text{in} \big | \big [ \hat{P}^{(e^-)}_s \psi(t, \mathbf{x} + \mathbf{s}/2), \, \overline{\hat{P}^{(e^-)}_s \psi (t, \mathbf{x} - \mathbf{s}/2) } \big ] \big | 0, \text{in} \big \rangle \,. 
\end{equation*}
To reexpress the result in terms of $W (\mathbf{p}, t)$, one should change the variable $\mathbf{x}'$ in Eq.~\eqref{eq:Ppsi} via $\mathbf{x}' = \mathbf{x}_0 + \mathbf{s}'/2$ together with $\mathbf{x}'' = \mathbf{x}_0 - \mathbf{s}'/2$ in the analogous integral involving the field-operator conjugate. Given the spatial homogeneity of the Wigner function, the integrals over $\mathbf{x}_0$, $\mathbf{p}'$, $\mathbf{p}''$, and $\mathbf{s}$ are easily computed, so we obtain
\begin{equation}
\hat{\mathcal{P}}^{(e^-)}_s W (\mathbf{p}, t) = u_{\mathbf{p},s} \overline{u}_{\mathbf{p},s} \gamma^0 W (\mathbf{p}, t) \gamma^0 u_{\mathbf{p},s} \overline{u}_{\mathbf{p},s} \,.
\end{equation}
The final expression turns out to be $\mathbf{x}$-independent as it should be. It is equivalent to
\begin{equation}
\hat{\mathcal{P}}^{(e^-)}_s \tilde{W} (\mathbf{p}, t) = u_{\mathbf{q},s} \overline{u}_{\mathbf{q},s} \gamma^0 \tilde{W} (\mathbf{p}, t) \gamma^0 u_{\mathbf{q},s} \overline{u}_{\mathbf{q},s} \,.
\end{equation}
According to Eq.~\eqref{eq:fs_proj}, the electron number density reads
\begin{eqnarray}
f^{(e^-)}_s(\mathbf{p}, t) &=& \mathrm{Tr} \ \bigg \{ \frac{m \mathrm{I} + \boldsymbol{\gamma} \mathbf{q}}{q^0} \, u_{\mathbf{q},s} \overline{u}_{\mathbf{q},s} \gamma^0 \big [ \tilde{W} (\mathbf{p}, t) - \tilde{W}_{\mathbf{A}=\mathbf{0}} (\mathbf{p}, t) \big ] \gamma^0 u_{\mathbf{q},s} \overline{u}_{\mathbf{q},s} \bigg \} \nonumber \\
{}&=& \frac{1}{q^0} \, \mathrm{Tr} \ \Big \{ \big ( \boldsymbol{\alpha} \mathbf{q} + \beta m \big ) u_{\mathbf{q},s} u^\dagger_{\mathbf{q},s} \big [ \tilde{W} (\mathbf{p}, t) - \tilde{W}_{\mathbf{A}=\mathbf{0}} (\mathbf{p}, t) \big ] \gamma^0 u_{\mathbf{q},s} u^\dagger_{\mathbf{q},s} \Big \} \nonumber \\
{}&=& \mathrm{Tr} \ \Big \{ \gamma^0 u_{\mathbf{q},s} u^\dagger_{\mathbf{q},s} \big [ \tilde{W} (\mathbf{p}, t) - \tilde{W}_{\mathbf{A}=\mathbf{0}} (\mathbf{p}, t) \big ]  \Big \} \,,
\label{eq:fs_app}
\end{eqnarray}
where we have used Eq.~\eqref{eq:bispinors_eqs_u} and the orthonormality of the bispinors. This result coincides with Eq.~\eqref{eq:fs_uud}.

\end{widetext}

%%%%%%%%%%%%%%%%%%%%%%%%%%%%%%%%%%%%%%%%%%%%%%%%%%%%%%%%%

\end{document}